\documentclass[sigconf]{acmart}
\AtBeginDocument{%
  }

\copyrightyear{2026}
\acmYear{2026}
\setcopyright{cc}
\setcctype{by}
\acmConference[CHI '26]{Proceedings of the 2026 CHI Conference on Human Factors in Computing Systems}{April 13--17, 2026}{Barcelona, Spain}
\acmBooktitle{Proceedings of the 2026 CHI Conference on Human Factors in Computing Systems (CHI '26), April 13--17, 2026, Barcelona, Spain}
\acmDOI{10.1145/3772318.3791615}
\acmISBN{979-8-4007-2278-3/2026/04}

\usepackage{amsmath}
\usepackage{booktabs}
\usepackage{float}
\usepackage{enumitem}

\makeatletter
\ifcase\ACM@format@nr
\relax % manuscript
\or % acmsmall
  \def\@authorfont{\large\sffamily}
  \def\@affiliationfont{\small\normalfont}
\or % acmlarge
\or % acmtog
  \def\@authorfont{\LARGE\sffamily}
  \def\@affiliationfont{\large}
\or % sigconf
  \def\@authorfont{\LARGE}
  \def\@affiliationfont{\small}
\or % siggraph
  \def\@authorfont{\normalsize\normalfont}
  \def\@affiliationfont{\normalsize\normalfont}
\or % sigplan
  \def\@authorfont{\Large\normalfont}
  \def\@affiliationfont{\normalsize\normalfont}
\or % sigchi
  \def\@authorfont{\bfseries}
  \def\@affiliationfont{\mdseries}
\or % sigchi-a
  \def\@authorfont{\bfseries}
  \def\@affiliationfont{\mdseries}
\or % acmengage
  \def\@authorfont{\LARGE}
  \def\@affiliationfont{\large}
\or % acmcp
  \def\@authorfont{\large\sffamily}
  \def\@affiliationfont{\small\normalfont}
\fi
\makeatother

\begin{document}

\title[Reflection-to-Action for Change]{Breaking Negative Cycles: \\ A Reflection-To-Action System For Adaptive Change}

\author{Minsol Michelle Kim}
\affiliation{%
  \institution{MIT Media Lab \\ Massachusetts Institute of Technology}
  \city{Cambridge}
  \state{Massachusetts}
  \country{USA}
}
\email{minsol@mit.edu}

\author{Daniel M. Low}
\affiliation{%
  \institution{Child Mind Institute}
  \city{New York}
  \state{New York}
  \country{USA}
}
\affiliation{%
  \institution{Harvard University}
  \city{Cambridge}
  \state{Massachusetts}
  \country{USA}
}
\email{daniel.low@childmind.org}

\author{David Lafond}
\affiliation{%
  \institution{Massachusetts Institute of Technology}
  \city{Cambridge}
  \state{Massachusetts}
  \country{United States}
}
\email{davidlaf@mit.edu}

\author{Eugene Shim}
\affiliation{%
  \institution{Wellesley College}
  \city{Wellesley}
  \state{Massachusetts}
  \country{United States}
}
\email{es118@wellesley.edu}

\author{Michelle Han}
\affiliation{%
  \institution{Massachusetts Institute of Technology}
  \city{Cambridge}
  \state{Massachusetts}
  \country{United States}
}
\email{mhhan@mit.edu}

\author{Mohanad Kandil}
\affiliation{%
  \institution{Technical University of Munich}
  \city{Heilbronn}
  \country{Germany}
}
\email{mohanad.kandil@tum.de}

\author{Chenyu Zhang}
\affiliation{%
  \institution{Harvard University}
  \city{Boston}
  \state{Massachusetts}
  \country{United States}
}
\email{chenyu_zhang@alumni.harvard.edu}

\author{Theo Kitsberg}
\affiliation{%
  \institution{University of Cambridge}
  \city{Cambridge}
  \country{United Kingdom}
}
\email{tck32@cam.ac.uk}

\author{Chelsea Boccagno}
\affiliation{%
  \institution{Harvard T.H. Chan School of Public Health}
  \city{Boston}
  \state{Massachusetts}
  \country{USA}
}
\email{cboccagno@g.harvard.edu}
\affiliation{%
  \institution{Massachusetts General Hospital}
  \city{Boston}
  \state{Massachusetts}
  \country{USA}
}

\author{Paul Pu Liang}
\affiliation{%
  \institution{MIT Media Lab, \\ Massachusetts Institute of Technology}
  \city{Cambridge}
  \state{Massachusetts}
  \country{USA}
}
\email{ppliang@mit.edu}

\author{Pattie Maes}
\affiliation{%
  \institution{MIT Media Lab, \\ Massachusetts Institute of Technology}
  \city{Cambridge}
  \state{Massachusetts}
  \country{USA}
}
\email{pattie@media.mit.edu}

% \author{Minsol Michelle Kim}
% % \authornote{Both authors contributed equally to this research.}
% \email{minsol@mit.edu}
% \orcid{}
% \author{}
% \affiliation{%
%   \institution{Massachusetts Institute of Technology}
%   \city{Cambridge}
%   \state{MA}
%   \country{United States}
% }

\renewcommand{\shortauthors}{Kim et al.}

\begin{abstract}
Breaking negative mental health cycles, including rumination and recurring regrets, requires reflection that translates awareness into behavioral change. Grounded in the Transtheoretical Model (TTM) and Gross’s Emotion Regulation (ER) Process Model, we examine how Technologies Supporting Self-Reflection (TSR) bridge reflection and action. In a 15-day in-the-wild study ($N = 20$), participants used a voice-based journaling system to capture regrets and wishes and engaged in \textbf{WhatIf-Planning}, a novel structured reflection module integrating counterfactual thinking with if--then planning. Participants were randomized to either a free-form condition or a Gross-guided condition, which maps the five processes of Gross’s ER model into explicit journaling prompts. We contribute: (1) a \textbf{unified reflection-to-action TSR system} that operationalizes the Preparation stage of TTM to bridge Contemplation and Action, and (2) triangulated empirical evidence from an in-the-wild journaling study that first operationalizes Gross’s Process Model, revealing effects on \textit{coping flexibility} and \textit{emotion regulation} in daily life. Results show significant pre--post improvements in coping flexibility, indicating adaptive self-regulation across conditions, with the Gross-guided group generating more counterfactual alternatives, articulating concrete if--then action plans, and implementing more plans for self-driven change.

\end{abstract}




% \ccsdesc[500]{Human-centered computing~User studies}  
% \ccsdesc[500]{Human-centered computing~HCI design and evaluation methods}  
% \ccsdesc[300]{Human-centered computing~Ubiquitous and mobile computing design and evaluation methods}  
% \ccsdesc[300]{Human-centered computing~Interaction techniques}  


\begin{teaserfigure}
  \centering
  \includegraphics[width=0.95\linewidth]{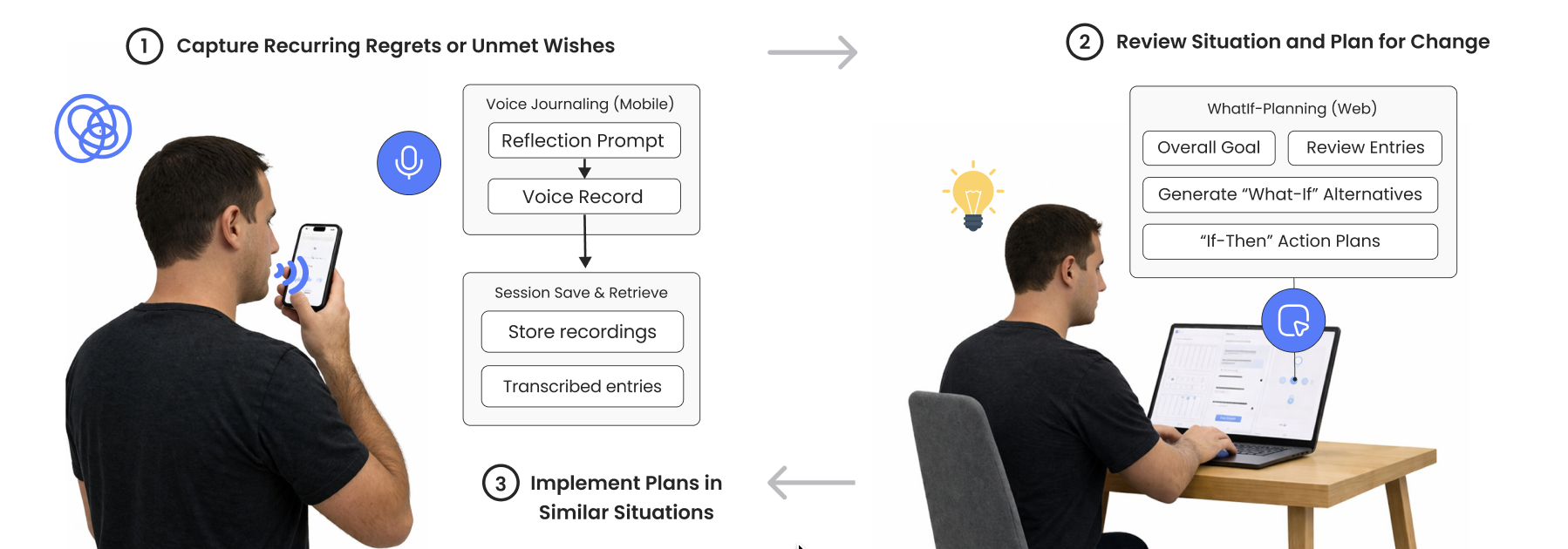}
  \vspace{-5pt}
  \caption[]{Overview of the Reflection-to-Action system. (1) The Voice Journaling mobile platform captures daily regrets or wishes through brief voice entries, which are automatically transcribed and stored. (2) In the WhatIf-Planning web platform, users generate counterfactual   ``what-if'' alternatives and create ``if-then'' action plans grounded in their recorded real-life scenarios. In the Gross-guided condition, we provide additional scaffolds to help users decode their patterns and strategize based on Gross’s Emotion Regulation Process Model. Together, these components support a reflection-to-action flow—from capturing lived experiences to preparing concrete strategies to implement for mindful behavior change.}
  \Description{Diagram showing the mobile and web applications of the reflection system.}
  \label{fig:overall-system}
\end{teaserfigure}

\maketitle
% \begin{document}

\section{Introduction}
Many people find themselves caught in repetitive negative cycles--replaying tense interactions, regretting past choices, or ruminating about what they “should have said,” without arriving at a clear path for change. Rumination, a form of \textbf{repetitive negative thinking}, involves dwelling on distress rather than problem-solving, amplifying negative affect~\cite{Watkins2020Rumination, NolenHoeksema2008Rethinking}. As a transdiagnostic risk and maintenance factor for psychopathology (e.g., depression, anxiety)~\cite{Ehring2008RNT, Michl2013Rumination}, chronic rumination undermines coping by keeping individuals stuck in unproductive loops.

Yet many existing technologies are not designed to disrupt these cycles. Existing digital well-being tools may facilitate logging experience, but rarely help users understand why patterns occur or how to respond. 
Effective self-regulation requires monitoring discrepancies and adjusting behavior accordingly~\cite{CarverScheier1982,CarverScheier1998}, yet most systems focus on tracking or nudging, rather than enabling adaptive adjustment. Without translating insight into action, reflection may fall into rumination. \textit{How can digital well-being tools help people move from awareness to adaptive change?}

To address this, we design and evaluate a digital well-being tool that integrates \textbf{counterfactual thinking}~\cite{Roese1997CounterfactualThinking, Bonanno2013RegulatoryFlexibility}, supporting the generation of “what-if” alternatives that imagine how past events could have unfolded differently. According to the functional theory of counterfactual thinking, counterfactuals shape behavioral intentions through a content-specific pathway~\cite{Epstude2008Counterfactual}. Prior work also demonstrates that structured reflection can transform raw emotions into coherent meaning~\cite{Trapnell1999reflection} and reduce reliance on maladaptive strategies such as rumination~\cite{Lyubomirsky1998rumination}. Yet little empirical work examines how to design systems that support counterfactual thinking for self-driven behavior change.  To guide system design, we draw on two complementary theories. The \textbf{Transtheoretical Model (TTM; \S\ref{sec:theory-TTM})} highlights the \textit{Preparation} stage as the bridge between contemplation (e.g., \textit{realizing that a small disagreement with a friend often escalates}) and action (e.g., \textit{deciding to pause and cool down before responding})~\cite{Prochaska1997transtheoretical}. \textbf{Gross’s Emotion Regulation (ER) Process Model} (\S\ref{sec:theory-Gross}) identifies multiple leverage points where reinterpretation and regulatory strategies (e.g., \textit{pausing to breathe or seeking the other person’s perspective}) can shift emotional trajectories~\cite{Gross2015extended}. Together, these models suggest opportunities for technology to scaffold progress from insight toward intentional behavior. 

% We introduce \textbf{WhatIf-Planning}, a \emph{Reflection-to-Action framework} that pairs mobile voice journaling with \textbf{what-if} counterfactual reflection and if–then implementation-intention planning (Figure~\ref{fig:overall-system}). The mobile app captures in-situ regrets and wishes, while the web interface guides users to generate \textbf{what-if} alternatives from recorded scenarios and translate them into actionable \textbf{if–then plans}.

% We introduce \textbf{WhatIf-Planning}, a \emph{Reflection-to-Action framework} pairing mobile voice journaling with \textbf{what-if} counterfactual reflection and if–then planning (Figure~\ref{fig:overall-system}), enabling the reframing of in-situ experiences into actionable plans. We investigate the following research questions through a 15-day in-the-wild study ($N=20$) under Free-form and Gross-guided conditions:

We introduce \textbf{WhatIf-Planning}, a \emph{Reflection-to-Action framework} pairing mobile voice journaling with \textbf{what-if} counterfactual reflection and if–then planning (Figure~\ref{fig:overall-system}), enabling the reframing of in-situ experiences into actionable plans. Deploying the system in an in-the-wild study ($N=20$), we investigate:
\vspace{-2pt}

\begin{itemize} \label{item:RQs}
    \item \textcolor{black}{\textbf{RQ1: Progression Toward Action.} How does a 15-day Reflection-to-Action intervention using voice journaling and counterfactual WhatIf-Planning (\S\ref{sec:system-overview}) support readiness for action, plan enactment, and coping flexibility?}
    
    \item \textbf{RQ2: Reflection Structure.} Does the \textit{Gross-guided condition} (\S\ref{sec:conditions}), a reflection scaffold grounded in Gross’s process model (\S\ref{sec:theory-Gross}), \textcolor{black}{differentially shape participants’ emotion-regulation processes and planning behaviors}?
    
    \item \textbf{RQ3: Voice Journaling Experience.} How do participants perceive and engage with \textit{voice-based journaling}? How do experiences differ by condition (\S\ref{sec:conditions})?
\end{itemize}
\vspace{-2pt}
% We address these questions through a 15-day in-the-wild study of the \textbf{WhatIf-Planning} system ($N=20$) under Free-form and Gross-guided conditions. 

This paper contribute: (1) a unified \textbf{reflection–to–action framework} integrating TTM’s Preparation stage with multi-stage emotion-regulation theory; (2) \textbf{the WhatIf-Planning module}, a modular system combining voice journaling, structured what-if reframing, and if–then planning through a novel Gross-guided scaffold; and 
(3) \textbf{empirical evidence from a 15-day in-the-wild study} demonstrating early improvements in Reflection-to-Action outcomes, such as coping flexibility, plan enactment, and emotion regulation.

\section{Theoretical Background} \label{sec:theory}
We outline three theoretical foundations for designing our reflection-to-action system. TTM and Gross’s Process Model of Emotion Regulation (ER) guide \textit{what} support is needed and \textit{when}, while the Mental Contrasting and Implementation Intentions (MCII) framework provides the mechanism through which reflection is translated into concrete action, motivating WhatIf-Planning module.

\vspace{-8pt}
\subsection{Transtheoretical Model (TTM) of Change} \label{sec:theory-TTM}
TTM conceptualizes behavior change as a staged process from inaction to sustained change~\cite{Prochaska1997transtheoretical}, through the cycle of:
\textit{(1) Precontemplation} (no intention to take action in the foreseeable future or unaware of problematic behavior);
\textit{(2) Contemplation} (recognize the problem and begin weighing the pros and cons of change);
\textit{(3) Preparation} (intend to take action in the immediate future and may begin taking small steps toward change);
\textit{(4) Action} (have made specific overt behavior changes); and
\textit{(5) Maintenance} (sustain behavior change and work to prevent \textit{Relapse}).
 Interventions are most effective when matched to an individual’s current stage, and progress through stages is supported by \textit{processes of change}, such as \textit{consciousness raising}, \textit{self-reevaluation}, and \textit{self-liberation}~\cite{Prochaska1997transtheoretical}. We target individuals in the \textbf{Contemplation} stage and design a system that explicitly scaffolds the \textbf{Preparation} stage to progress toward \textbf{Action}, a transition existing systems rarely support (\S\ref{sec:reflection-tech}).

\vspace{-5pt}
\subsection{Emotion Regulation (ER) Process Model}\label{sec:theory-Gross}

Gross's ER Process Model conceptualizes an emotional episode as a dynamic sequence of regulatory opportunities~\cite{Gross1998emotion, Gross2015extended}, comprising five families of strategies: 
\textit{(1) Situation selection} (approaching or avoiding situations to influence which emotions arise); 
\textit{(2) Situation modification} (directly altering features of a situation to change its emotional impact); 
\textit{(3) Attentional deployment} (directing or shifting attention within a situation); 
\textit{(4) Cognitive change} (reinterpreting the meaning of a situation to alter its emotional significance); and 
\textit{(5) Response modulation} (influencing experiential, behavioral, or physiological responses once an emotion is underway). 

These strategies highlight that emotion regulation unfolds across multiple leverage points before, during, and after an emotional response. Together with TTM, this framework suggests that effective systems should \textit{support users in recognizing where they are within the regulatory sequence and adopting contextually appropriate strategies}, rather than providing one-size-fits-all support. This conceptualization aligns with contemporary theories of regulatory flexibility and context-sensitive emotion regulation~\cite{Troy2021Emotion, Dore2016Personalized}. In our Gross-guided condition~(\S\ref{sec:conditions}), we provide prompts aligned with each strategy, enabling individuals to reflect, review, and plan across multiple regulatory opportunity points to expand their coping repertoire.

\vspace{-5pt}
\subsection{Mental Contrasting and Implementation Intentions (MCII)} ~\label{sec:theory-IIMC}

MCII is a validated self-regulation framework for bridging intentions and actions~\cite{Sefidgar2024IIMC}. \textit{Mental Contrasting (MC)} guides individuals contrast desired futures with present obstacles, strengthening goal commitment when goals are appraised as feasible~\cite{oettingen2009mental} and increasing energization, a marker of motivational readiness~\cite{Grant2012WorkMotivation}. \textit{Implementation Intentions (II)} translate this commitment into concrete “if–then’’ plans linking anticipated obstacles to contingent coping responses~\cite{Gollwitzer1999implementation}, enabling automatic initiation of goal-directed actions while reducing real-time cognitive load.

\textbf{The WOOP model} (Wish, Outcome, Obstacle, Plan) is an example that operationalizes MCII into a simple, structured practice and has been shown to improve self-control and goal attainment across academic, social, and health domains~\cite{stadler2010intervention, duckworth2011self, krott2018mental}. However, WOOP is inherently future-oriented, relying on imagined wishes and obstacles rather than recent lived experience. Consequently, this model cannot help users reinterpret past events or uncover situational drivers of struggle (e.g., \textit{making an impulsive purchase after sudden stress.}) To address this limitation, we introduce the \textbf{WhatIf-Planning} framework, which replaces WOOP’s imaginative “Wish–Outcome’’ steps with \textit{what-if counterfactuals} grounded in real events and distills them into \textit{actionable if–then plans}, aligned with WOOP’s “Obstacle–Plan’’ components.

\vspace{-5pt}
\section{Related Work}

\subsection{Technology-Supported Reflection (TSR) }~\label{sec:reflection-tech}

Human-Computer Interaction (HCI) has increasingly explored how wearable sensing, adaptive prompts, and large language models (LLMs) can support self-reflection. AI-enhanced journaling systems such as \textit{MindScape}~\cite{Nepal2024MindScape} and \textit{MindfulDiary}~\cite{Kim2024MindfulDiary} show that context-aware prompts and empathetic dialogue can deepen emotional disclosure and sense-making. \textit{ExploreSelf}~\cite{Song2025ExploreSelf} similarly uses LLM-driven adaptive questioning to help participants reflect on personal challenges and self-selected themes, offering flexible and personalized narrative support. However, these systems focus on momentary insight generation without examining how such insights are carried forward or applied when similar situations recur in daily life. \textit{CounterStress}~\cite{Jung2025CounterStress} builds a personalized stress-prediction model from contextual and self-reported data to generate counterfactual, lower-stress alternatives. However, its support remains focused on situational suggestions without supporting users in examining longer-term strategy or capacity development.

This pattern echoes Hao et al.’s systematic review of \emph{Technology-Supported Reflection (TSR)}, which found that most of the 23 reviewed systems for self-reflection \textcolor{black}{on social interactions} focus on the \emph{review and articulation} of past experiences--typically through summaries, visualizations, or guided prompts~\cite{Hao2025TSRReview}. Related work on reflection inventories~\cite{Bentvelzen2021tsri}, and design resources for reflective technologies~\cite{Slovak2022reflection} \textcolor{black}{similarly highlights that many systems facilitate basic sense-making but rarely skill-development. Existing HCI systems have rarely examined how reflective scaffolds might align with, or support progression through, TTM.}

\vspace{-5pt}
\subsection{Translating Reflection into Action}\label{sec:psych-frameworks}
Outside HCI, psychological research provides empirically evaluated frameworks for bridging insight and behavior. As introduced in \S\ref{sec:theory-IIMC}, implementation intentions (“if–then” plans) reliably increase goal enactment across domains~\cite{Gollwitzer1999implementation}, and MCII—often operationalized through the WOOP sequence—has been validated across academic, interpersonal, and health contexts~\cite{stadler2010intervention,duckworth2011self,krott2018mental}.

Recent HCI work has incorporated MCII-based planning into digital tools; for example, Sefidgar et al.~\cite{Sefidgar2024IIMC} applied MCII in a workplace system and improved goal clarity and metacognitive awareness. Yet journaling and reflection interfaces seldom integrate such mechanisms, often stopping at emotional disclosure or pattern recognition. Little is known about how reflective insights translate into structured preparation or everyday action, or how they align with psychosocial models such as TTM. We address this gap by investigating a Reflection-to-Action framework that strengthens regulatory capacities and supports adaptive trajectory change in line with Gross’s ER model and TTM.

\subsection{Voice-Based Reflection for Self-Regulation}\label{sec:voice-reflection}

Voice-based interaction offers several affordances for reflection-support technologies. Prior work shows that voice-based journaling often produces longer and more open reflections than text-based, suggesting that speaking out loud may encourage more spontaneous expression~\cite{Sayis2024hri_journaling}. \textcolor{black}{ Similarly, audio-diary research~\cite{Crozier2016audioDiaries} showed that spoken entries capture experiences closer to the moment they occur and yield more spontaneous, affectively rich accounts than written logs.} A recent review of self-disclosure to conversational AI~\cite{Papneja2025selfDisclosure} identifies \textit{interface modality} (e.g., voice vs. text) as a key factor shaping user openness, with spoken interaction often associated with more fluid and spontaneous forms of disclosure and richer affective expression, when psychological safety is maintained.

Together, these findings suggest that voice journaling supports deeper, more natural self-reflection, especially when immediacy, affective nuance, and a non-judgmental space matter. However, most voice-based systems emphasize expressive disclosure or scripted guidance and do not examine structured cognitive reinterpretation or daily reflection as pathways to self-regulation. We address this gap by pairing voice journaling with the WhatIf-Planning module and evaluating user experiences in a 15-day in-the-wild study.

\section{Methodology}
We conducted a 15-day mixed-methods study examining how daily voice journaling and weekly WhatIf-Planning support progression toward action \textbf{(RQ1)}, how reflection structure (Gross-guided vs. Free-form; \S\ref{sec:conditions}) shapes adaptive coping and emotion regulation \textbf{(RQ2)}, and how participants engage with voice journaling \textbf{(RQ3)}.

\subsection{System Overview and Design Rationale}\label{sec:system-overview}

Building on modular perspectives in psychosocial technology design~\cite{Slovak2024framework}, our system comprises two core modules: \textbf{daily voice journaling} and \textbf{WhatIf-Planning}. The mobile application captures in-situ spoken reflections, while the web interface supports counterfactual “what-if’’ review and weekly "if–then planning" \textcolor{black}{(Figure~\ref{fig:overall-system}; \S\ref{sec:system-design}). }\textcolor{black}{Together, these modules form an \textbf{integrated Reflection-to-Action platform} that operationalizes the frameworks in \S\ref{sec:theory}: they scaffold intermediate \textit{readiness-to-act} (\S\ref{sec:measures-SystemSupport}) within the Preparation stage and help users progress from Contemplation toward Action in TTM (\S\ref{sec:theory-TTM}), while translating reflective insight into concrete, context-sensitive plans through MCII mechanisms (\S\ref{sec:theory-IIMC}). To examine how guided structure shapes this process, we compare a \textbf{Gross-guided} scaffolding based on the Gross ER process model (\S\ref{sec:theory-Gross}) with \textbf{Free-form} baseline, outlined below.}

\subsection{Conditions.}~\label{sec:conditions} We conducted a 15-day study in which all participants completed the same journaling and What-If-Planning tasks under one of two conditions: 
\begin{itemize}
    \item \textbf{Free-form (N=10).} Participants are prompted with a single minimal guidance prompt \textcolor{black}{(Table~\ref{tab:prompts-free-form})} designed to support natural, unguided reflection. \textcolor{black}{This condition served as a baseline for understanding how individuals naturally reflect, identify alternative opportunities for change, and construct If–Then implementation intentions.}
    \item \textbf{Gross-guided (N=10).} Participants used prompts (Table~\ref{tab:prompts-grossGuided}) aligned with the five stages of Gross’s ER Process Model. These prompts were designed to deepen emotional reflection by scaffolding reflection at multiple leverage points in the regulation process and helping with translation of insights into actionable plans. Participants could skip prompts if a prompt does not apply to their situation.
\end{itemize}

\textcolor{black}{This condition setup enabled us to examine the benefits of stage-based scaffolding for emotion regulation (RQ2) and its added value in supporting readiness to change beyond the What-If-Planning module alone (RQ1). We also examined whether participants’ perceptions and experiences with voice journaling differed by condition (RQ3). Condition-specific prompts (Appendix~\ref{appendix:prompts-full}) were delivered across the mobile and web interfaces.}

\vspace{-8pt}
\subsection{Participants and Recruitment Strategy}~\label{sec:recruitment}
\textcolor{black}{We randomly selected twenty participants from a pool of 100 respondents (e.g., students, staff, and affiliates) recruited via departmental and dormitory mailing lists at institutions in Massachusetts and stratified them into two conditions.} Eligibility was determined through a two-stage screening process: (1) a \textbf{participant screening survey} with general inclusion criteria (e.g., at least 18 years old, English fluency, comfort with verbal reflection) and exclusion criteria for individuals who reported conditions that might interfere with participation (e.g., certain psychiatric diagnoses); and (2) a \textbf{pre-survey} \textcolor{black}{to identify participants with recurring regrets or unmet wishes, consistent with the \textit{Contemplation stage}, and to assess feasibility for progress within the 15-day study period.} Upon completing the study, participants received a \$60–\$80 Amazon gift card, based on their interview and weekly session duration.

% This was to target individuals in the \textbf{Contemplation stage} of TTM, or those who recognize a recurring issue but have not yet taken consistent action to address it.
% \paragraph{Pre-survey and Scenario Eligibility.}

In the initial pre-survey (Appendix ~\ref{appendix:pre-survey-items} for full items), respondents \textbf{(N=100)} listed up to three recurring regret or unmet wish scenarios across relational and non-relational domains that were expected to recur multiple times per week during the study window. Respondents rated scenarios on a 5-point Likert scale across:
\begin{itemize}
    \item \textbf{Opportunity for Change} ($\geq 3$): at least moderately controllable by the participant, rather than primarily driven by external constraints, ensuring feasibility for meaningful progress during the 15-day period (RQ1).
    \item \textbf{Emotional Intensity} ($\geq 3$): at least a moderate level of negative emotion or regret, capturing scenarios with sufficient affective salience to observe regulation effects (RQ2, RQ3).
    \item \textbf{Personal Importance} ($\neq 1$): excluded scenarios rated unimportant, to ensure sufficient motivation and align with the TTM Contemplation stage (RQ1).
\end{itemize}
\vspace{-2pt}
Participants who submitted scenarios received \$$1$ per scenario regardless of eligibility. Pre-survey data were also used to characterize the broader landscape of recurring regrets and wishes (Appendix~\ref{Appendix:InsightsFromTargetScenarios}). 

\begin{table}[t]
    \caption[]{Participant ($N=20$) demographics and background. Values indicate counts. Age and gender were collected during recruitment, while prior journaling and recent voice-journaling experience were collected in the baseline questionnaire after recruitment. Gender was used to balance participants across conditions.}

  % \caption[]{Participant ($N=20$) demographics and background. Values represent counts. Prior journaling or voice-use experience indicates the number of participants reporting previous experience with these practices. \textcolor{black}{Age and gender were collected during the recruitment stage, while prior journaling and recent voice-journaling experience were reported in the baseline questionnaire after recruitment. Gender was used to split participants into two condition groups.}}
  \label{tab:demographics}
  \small
  \begin{tabular}{lcc}
    \toprule
    \textbf{Condition Group Background} & \textbf{Free-Form} & \textbf{Gross-Guided} \\
    \midrule
    Number of Participants            & 10 & 10 \\
    Age (mean, SD)          & 25.4 (11.06) & 26 (5.12) \\
    Gender (F/M/Other)      & 8 / 2 / 0  & 7 / 2 / 1 \\
    % Recent journaling (Yes)  & 7 /20 & 8 /20 \\
    \textcolor{black}{Prior journaling (Yes)}  &    \textcolor{black}{20 /20} &    \textcolor{black}{20 /20}\\
    Recent voice use in journaling (Yes)   & 0 /20 & 0 / 20 \\
    \bottomrule
  \end{tabular}
\end{table}

\paragraph{Final Focus Group.}  \textcolor{black}{Twenty respondents from the non-clinical population who, after the screening survey, identified at least one eligible scenario in the pre-survey were invited to participate in the study as the final focus group} (Table~\ref{tab:demographics}). Participants were stratified into the two study conditions while balancing scenario type and gender. Although age was not explicitly balanced, the groups' average ages were comparable. When individuals listed multiple qualifying scenarios, one scenario was automatically selected to maintain a balanced representation of relational and non-relational themes (Appendix~\ref{appendix:pre-survey} for full details) between conditions. \textcolor{black}{To preserve privacy, the research team did not directly access or review the content of participants’ identified scenarios. Instead, participants were asked to self-categorize each scenario as relational or non-relational in the pre-survey.}

\vspace{-3pt}
\begin{figure*}[t]
 \centering
 \includegraphics[width=0.85\linewidth]{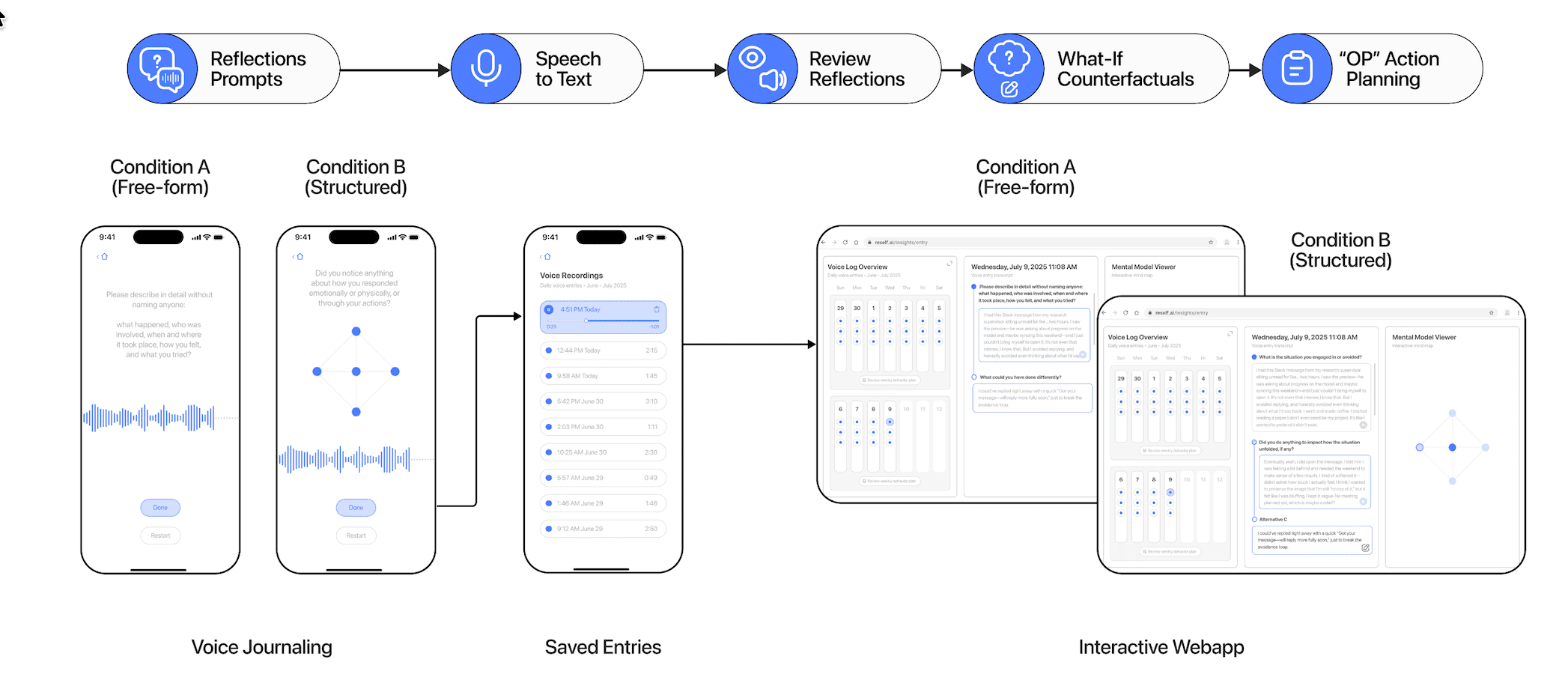}
 \caption[]{Overview of the study interfaces. Participants journaled using a mobile application. Entries were synchronized to a web interface, where participants reviewed transcribed journal entries, tracked progress via a calendar, and engaged in WhatIf-Planning by generating “what-if” counterfactual alternatives and translating them into implementation-intention plans.}

 \label{fig:SystemDiagram}
 \end{figure*}

\begin{figure*}[t]
\centering
\includegraphics[width=0.9\linewidth]{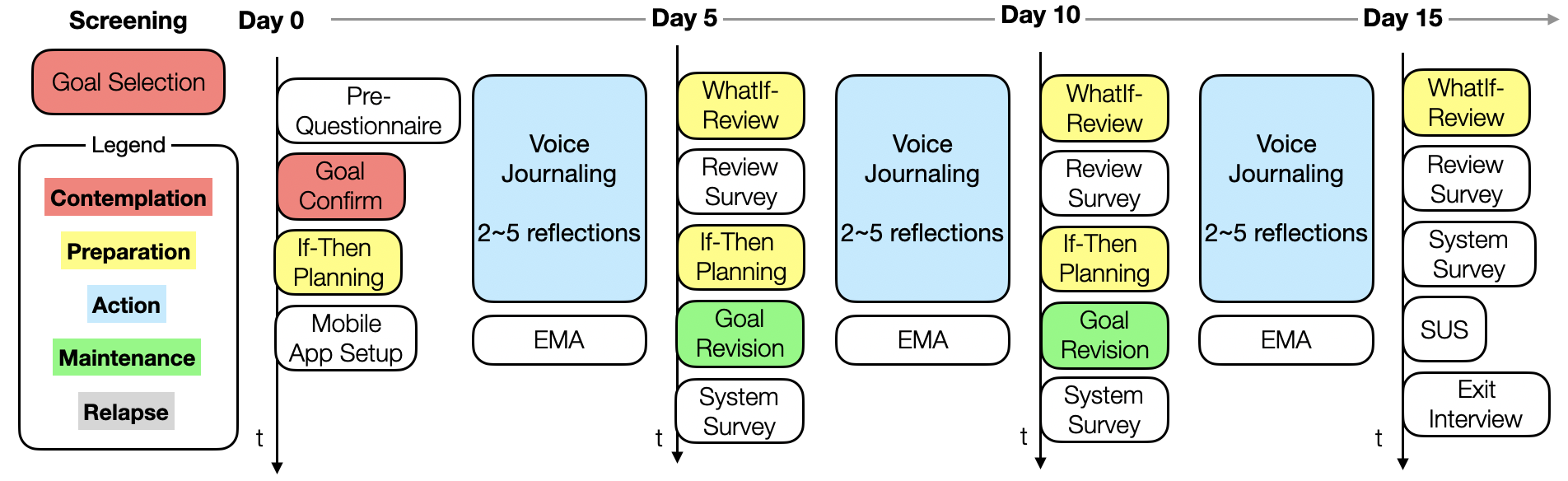}
\caption[]{Overview of the study procedure. The horizontal axis shows the timeline (Days~0, 5, 10, 15), and the vertical axis shows tasks within each facilitated session. Colored labels indicate alignment with Prochaska’s Transtheoretical Model (TTM). Participants completed 2–5 voice reflections per 5-day interval at their discretion.}

% \caption[]{Overview of 15-day procedure. Horizontal flow shows timeline (Days~0, 5, 10, 15). Vertical flow shows tasks within each facilitated session. Colored labels indicate relevance to stages of Prochaska’s Transtheoretical Model (TTM). Voice journaling occurred at participants’ discretion between each 5-day interval, with participants completing 2–5 reflections per interval.}
\label{fig:study-procedure}
\end{figure*}

\subsection{Study Procedure}\label{sec:study-procedure}
The study spanned 15 days, divided into three 5-day intervals (Fig~\ref{fig:study-procedure}). This duration gave participants enough time to naturally re-encounter their recurring scenarios in daily life while balancing study costs and participant burden. Additionally, the three intervals enabled repeated measurements on Days~5, 10, and 15, allowing us to track within-subject change over the course of the 15-day intervention. Conditions differed only in the prompts used during daily journaling and weekly planning (see Appendix~\ref{appendix:prompts-full}). All participants completed the following tasks:

\begin{enumerate}
    \item \textbf{Onboarding and Baseline (Day~0; in-person).} 
    Completed the consent form, baseline questionnaires, and confirmed the overall goal related to their recurring wish or regret scenario that they would work on for the next 15 days. In addition, participants authored an initial If–Then plan for the first interval (Days~1–5). 

    \item \textbf{Daily Voice Journaling (Days~1–15; in-the-wild).}
    Recorded in-situ voice journals whenever relevant events occurred that they wished to revisit for their overall goal. They were encouraged to record at least two entries per 5-day interval (Days 1-5, 6-10, 7-15).

    \item \textbf{WhatIf-Planning Sessions (Days~5, 10, 15; remote via Zoom).}
    \textcolor{black}{Attended brief remote sessions with a randomly assigned trained co-author facilitator (\S\ref{sec:facilitators}). Each session followed a standardized two-part structure, while all reflective work was completed independently by participants:}
    \begin{itemize}
        \item\textcolor{black}{\textbf{(a) WhatIf-Review (2–4 minutes per entry).}}
        \textcolor{black}{Reviewed their transcribed entries, generated counterfactual alternatives (\S\ref{sec:measures-weeklyInteraction}), and completed the \textit{Weekly WhatIf Review} survey (\S\ref{sec:measures-WhatIfReview}).} 
        \item \textcolor{black}{\textbf{(b) If–Then Planning (up to 20 minutes).}
        Generated If-then plans (\S\ref{sec:measures-weeklyInteraction}) to experiment and implement for the next interval.}
    \end{itemize}

    \textcolor{black}{At the end of each WhatIf-Planning session, facilitators administered the \textit{System Support for Readiness-to-Action} survey (\S\ref{sec:measures-SystemSupport}). Participants had an option to disable audio/video during the session to reduce distraction.}

    \item \textbf{Post-Study Questionnaire and Exit Interview (Day~15; remote).}
    Completed a final questionnaire, participated in a semi-structured interview, and received information about reimbursement.
\end{enumerate}
\vspace{-5pt}

\textcolor{black}{\paragraph{Facilitation and Protocol Consistency.}~\label{sec:facilitators}
Facilitators (the first author and three trained co-author facilitators) ensured consistent procedures during onboarding and offboarding and controlled session duration across participants during the weekly WhatIf-Planning sessions. \textbf{Onboarding} was facilitated in person, primarily to help with mobile application setup and to answer questions about app usage and the voice journaling activities participants would complete independently. Participants scheduled \textbf{Weekly WhatIf-Planning sessions} with the facilitators. During the session, the facilitators provided instructions, administered surveys, and controlled the review and planning durations to keep engagement time during each WhatIf-Planning activity comparable across participants. Aside from receiving instructions, all reflective tasks were completed independently, and participants were allowed to turn off audio and video while completing their weekly activities and surveys. \textbf{Post-Study Questionnaire and Exit Interview} on Day~15 included administering the post-study questionnaire, with the first author conducting the exit interview and a second facilitator serving as note taker. All facilitators were trained by the first author and followed a shared protocol to maintain consistency across sessions.}

\subsection{Study Interface}
\label{sec:system-design}
The 15-day study was supported by an integrated platform (Figure~\ref{fig:SystemDiagram}) consisting of a React/TypeScript web application, a React Native mobile client, and Firebase services for storage, authentication, and audio transcription (full system pipeline in Appendix~\ref{Appendix:Full-System}):
\vspace{-3pt}
\begin{itemize}
    \item \textbf{Voice Journaling (Mobile).} \textcolor{black}{Enabled journaling and was deployed through Expo Go~\cite{ExpoGo2024}. Participants recorded short scenario-related entries, which were locally cached on-device, uploaded when online, and automatically transcribed and stored in Firebase.}

    \item \textbf{WhatIf-Planning (Web).} \textcolor{black}{Provided synchronized access to recent entries during each session, enabling participants to: (i) review reflections and generate counterfactual "What if" alternative interpretations or actions, and (ii) translate these insights into concrete "If--Then" plans using the embedded implementation-intention planner. All interaction logs were stored in Firebase.}
\end{itemize}

Free-form participants received open-ended prompts (Table~\ref{tab:prompts-free-form}, whereas Gross-guided participants received prompts (Table~\ref{tab:prompts-grossGuided} aligned with the five stages of Gross’s ER Process Model. All other system behaviors and activities were identical across conditions.

\section{Measures}\label{sec:measures}

To evaluate how the \textbf{WhatIf-Planning} and \textbf{Voice Journaling} modules impacted adaptive coping, readiness for action, and reflective experience, we collected (i) validated pre–post psychological scales (\S\ref{sec:measures-CFS}, \S\ref{sec:measures-DERS}), (ii) daily and weekly in-situ self-reports (\S\ref{sec:measures-WhatIfReview}), (iii) interaction logs from mobile and web components (\S\ref{sec:measures-weeklyInteraction}, \S\ref{sec:measures-dailyInteraction}), and (iv) post-study semi-structured interviews (\S\ref{sec:measures-interview}). Together, these multi-level data sources triangulate both proximal shifts in reflective practice and broader changes in self-regulation across the 15-day deployment.

% \section{Measures}~\label{sec:measures}

% \subsection{Adaptive Coping and Readiness for Action (RQ1)}

\subsection{Readiness for Action and Coping (RQ1, RQ2)}

\subsubsection{Coping Flexibility Scale–Revised (CFS-R; Pre-Post).}~\label{sec:measures-CFS}
Because WhatIf-Planning prompts users to reconsider habitual responses, identify obstacles, and generate alternative interpretations, we included a measure that captures people’s ability to evaluate their current coping approach and adapt it when needed. The \textit{Coping Flexibility Scale–Revised (CFS-R)}~\cite{kato2012development, kato2020cfsr} directly assesses this capacity through three mechanisms: \textbf{Meta-coping} (monitoring and evaluating coping efforts), \textbf{Abandonment} (disengaging from ineffective strategies), and \textbf{Re-coping} (generating and applying alternative strategies). These components closely mirror the cognitive steps elicited by WhatIf-Planning—evaluating one’s typical response, recognizing when it is unhelpful, and generating alternative strategies—making the CFS-R theoretically aligned with the skills our intervention targets. Thus, we used total and subscale scores to evaluate adaptive coping \textbf{(RQ1)} and to understand whether scaffolded reflection differentially impacted these outcomes \textbf{(RQ2)}.

\vspace{5pt}
\subsubsection{WhatIf Review Session Items (Weekly).}~\label{sec:measures-WhatIfReview}
\textcolor{black}{We examined weekly self-report items during their WhatIf review session (Days~5, 10, and 15) to capture participants’ behavioral adoption of their plans and their perceived \textit{readiness for action}, by asking the participants to report how many of their planned actions they tried (\textit{Number of Plans Enacted}), how many of their previously identified obstacles occurred (\textit{Number of Obstacles Occurred}), and how effective and helpful participants found the planning components (\textit{Perceived Plan Effectiveness}, \textit{Set Goal Helpfulness}, and \textit{Obstacle Identification Helpfulness}). These questions are directly relevant to whether participants' intentions were translated into concrete behavioral adoption, probing the progression from preparation to action in TTM \textbf{(RQ1)}.}

\vspace{4pt}
\subsubsection{WhatIf-Planning System Logs (Weekly).}~\label{sec:measures-weeklyInteraction} 
To track observable actions that reflect counterfactual thinking and planning processes, the web system automatically saved participants' \textbf{What-If alternatives} generated and \textbf{obstacles identified} during each weekly planning session, which we quantified using simple counts. These behavioral traces complement both CFS-R change scores and weekly self-reports, enabling fine-grained analysis of concrete planning behaviors underlying readiness for action \textbf{(RQ1)} and potential differences between scaffolds \textbf{(RQ2)}.
\vspace{4pt}
\subsubsection{System Support Survey for Readiness-to-Action (Weekly).}~\label{sec:measures-SystemSupport}
To assess whether users formed \textit{concrete, situation-specific intentions} after engaging with WhatIf-Planning, we administered four weekly items on Days~5, 10, and 15. These items measured perceived support for obstacle identification, plan adjustment, short-term implementation intentions, and longer-term strategy reflection. Together, they capture how the WhatIf-Planning system supported \textit{readiness for action}, indexing participants’ weekly progression.

\vspace{5pt}
\subsection{Emotion Regulation  (RQ2, RQ3)}
\vspace{3pt}
\subsubsection{Difficulties in Emotion Regulation Scale (DERS; Pre-Post Measure).} ~\label{sec:measures-DERS}
To assess emotion-regulation capability, we administered the \textit{DERS-SF}~\cite{kaufman2016derssf}, which measures six domains of regulation difficulty: non-acceptance, goal-directed behavior, impulse control, emotional awareness, emotional clarity, and access to regulation strategies. The relevant subscales for our system are emotional clarity, a construct associated with reduced rumination~\cite{Thompson2017clarity} and directly targeted by the interpretive depth of the Gross-guided prompts, goal-directed behavior, and access to regulation strategies, which correspond to the cognitive and strategy-generation processes the stage-based scaffold is designed to support. Changes in these subdomains can indicate whether structured prompts improved regulatory insight beyond free-form reflection \textbf{(RQ2)}.

\vspace{5pt}
\subsubsection{Voice Journaling Interaction (Daily).}~\label{sec:measures-dailyInteraction}
To contextualize outcome differences, we analyzed aspects of daily journaling behavior--specifically, the number of entries, session duration, and cumulative reflection time--to characterize engagement with the voice modality \textbf{(RQ3)}. These indicators help determine whether differences in adaptive coping and readiness for action or emotion regulation \textbf{(RQ2)} is due to reflective “dose’’ or the Gross-guided scaffold. 

\subsection{System Engagement and Usability (RQ3)}~\label{sec:measures-engagement}

We analyzed daily journaling behavior—number of entries, session duration, and cumulative reflection time—to characterize engagement with the voice modality \textbf{(RQ3)}.

% \textcolor{black}{These indicators also help contextualize whether differences in adaptive coping, readiness for action, or emotion regulation reflect reflective “dose” or scaffolded structure; for example, similar reflection time with divergent outcomes would suggest condition-specific effects rather than time spent reflecting.}

\subsubsection{Voice Journaling System Engagement and System Usability Score (SUS; Post)}~\label{sec:measures-SUS}
Interaction logs were used to characterize how often and for how long participants engaged with the voice-journaling module. To assess overall usability and learnability, participants completed the \textit{System Usability Scale (SUS)}~\cite{lewis2018system}, a widely used 10-item questionnaire for evaluating perceived ease of use of a system.

\subsubsection{Semi-structured Interview (Post-study)}~\label{sec:measures-interview}
Finally, to capture participants’ subjective experiences with voice journaling \textbf{(RQ3)}, we conducted a brief semi-structured interview (20–40 minutes) on Day~15. Interviews focused on participants’ perception and experiences with daily voice journaling, the WhatIf-Planning module, and the Gross-guided prompts. All interviews were audio-recorded and transcribed for later qualitative analysis.

\section{Data Analysis}\label{sec:analysis}
\textcolor{black}{All participants across conditions (see \S\ref{sec:conditions}) completed daily mobile voice-journal entries, weekly WhatIf-Planning sessions to capture intermediate progression toward action (\S\ref{sec:measures-WhatIfReview}--\S\ref{sec:measures-SystemSupport}), and pre/post questionnaires (CFS--R, \S\ref{sec:measures-CFS}; DERS--SF, \S\ref{sec:measures-DERS}) for overall user outcomes. Across all analyses, we used two-tailed tests with $\alpha = .05$ and report effect sizes and confidence intervals.}

% \\(CFS--R for RQ1; DERS--SF for RQ2)
\vspace{-5pt}
\subsection{Pre--Post User Outcomes (RQ1, RQ2)}\label{sec:analysis-prepost}
To evaluate overall intervention effects on coping flexibility and emotion-regulation difficulties \textbf{(RQ1, RQ2)}, we conducted \textbf{2~(Phase: pre vs.\ post)~$\times$~2~(Condition: Free-form vs.\ Gross-guided) mixed ANOVAs} on \textit{CFS--R Total} and \textit{DERS--SF Total}, where phase was modeled as a within-participant factor and condition as a between-participant factor. Holm correction was applied across the two primary outcomes. 

As our sample size was underpowered for detecting interaction effects for all \textit{CFS-R and DERS-SF subscales}, we further complemented the overall ANOVAs with a targeted subscale-level \textit{Post--Pre} difference ($\Delta$) analysis to understand which regulatory mechanisms might be most responsive to the intervention and promising for future design. For each CFS--R and DERS--SF subscale, $\Delta$ scores were compared across conditions using \textbf{Welch’s $t$-tests} (robust to unequal variance), and primary investigated into \textbf{Hedges’ $g$} for effect-size estimate as significance tests on the subscales are underpowered for our pilot (Table~\ref{tab:delta_between_conditions}). These patterns are visualized in a forest plot showing subscale-level change and 95\% CIs (Figure~\ref{fig:subscale_forest}), providing a descriptive complement to the mixed ANOVA results.

\subsection{WhatIf-Planning Weekly Support for Readiness for Action (RQ1, RQ2)}\label{sec:analysis-weekly}
To understand how the \textbf{weekly WhatIf-Planning intervention} supported participants’ readiness for action \textbf{(RQ1)} and how this process differed by reflective scaffold \textbf{(RQ2)}, we analyzed two sets of weekly survey measures: (1) \textit{Weekly WhatIf Review Items} (\S\ref{sec:measures-WhatIfReview}) and (2) \textit{WhatIf-Planning System Support Items} (\S\ref{sec:measures-SystemSupport}).

As the weekly survey data were administered at three time points (on Days 5, 10, and 15), we used a linear mixed model to account for the repeated measures. Specifically, we used \textbf{piecewise linear mixed-effects models}, which allow the estimation of separate linear slopes for each interval, based on our hypothesis that change patterns might differ between the early phase of the study (Days~5--10) and the later phase (Days~10--15), as participants made progress and may have had fewer new plans to generate. Each model included random intercepts for participants and fixed effects for Condition (Free-form vs.\ Gross-guided), segmented time (seg1: Days~5--10; seg2: Days~10--15), and their interactions (Condition~$\times$~seg1; Condition~$\times$~seg2).  For comparison, we also fit single-slope models with Day as a continuous predictor and a Day~$\times$~Condition interaction.
Fixed-effect estimates (coefficients, $z$-values, and 95\% confidence intervals) from these models are used to interpret readiness-for-action (\S\ref{sec:results_readiness_for_action}) trajectories in the Results section.

% \subsection{Mobile and Web System Log during Voice Journaling and WhatIf-Planning (RQ1-RQ3)}

\subsection{System Interaction Log (RQ1–RQ3)}~\label{sec:analysis-system}
To objectively assess participants’ behavioral interaction patterns beyond self-report measures, we further analyzed \textit{mobile and web interaction logs} from daily voice journaling (\S\ref{sec:measures-dailyInteraction}) and weekly WhatIf-Planning activities (\S\ref{sec:measures-weeklyInteraction}). To evaluate condition differences, we again applied \textbf{Welch’s independent-samples $t$-tests} to each aggregated engagement metric. Effect sizes were estimated using Hedges’ $g$ and are summarized in Table~\ref{tab:behavioral-engagement}. This analysis also allowed us to contextualize voice journaling ``dose’’ across conditions to help interpret the psychological and planning outcomes reported in the previous subsections.

\subsection{Interivew Thematic Analysis (RQ1--RQ3).}~\label{sec: interviewAnalysis}
We analyzed interview transcripts using Braun and Clarke’s six-phase thematic analysis approach~\cite{BraunClarke2006}. Two researchers independently conducted hybrid coding that combined theory-driven categories with inductive codes, refined a shared codebook through continuous comparison, and resolved discrepancies through discussion. Codes were organized into candidate themes, reviewed for coherence, and finalized with analytic memos. All analysis was conducted in MAXQDA~\cite{MAXQDA}.

\section{Results}
We report quantitative findings on the collected data (\S\ref{sec:measures}) following the analysis structure in \S\ref{sec:analysis}. In \S\ref{sec:results_cfs_ders}, we first report pre--post psychological outcomes and subscale patterns (Fig.~\ref{fig:subscale_forest}, Tables~\ref{tab:anova_results_cfs_ders}--\ref{tab:delta_between_conditions}). Then, in \S\ref{sec:results_readiness_for_action}, we examine weekly self-report review and readiness-for-action system support survey as well as system log on weekly activities (Figs.~\ref{fig:weekly_whatif} --\ref{fig:weekly_system_support}, and Panels 2a-c of Fig.~\ref{fig:behavioral_engagement}). Finally, in \S\ref{sec:results_journaling_engagement}, we characterize voice journaling engagement (Panel 1a-1c of Fig.~\ref{fig:behavioral_engagement}).
\vspace{-5pt}

\subsection{Pre--Post Psychological Outcomes and Subscale Patterns (RQ1, RQ2)}\label{sec:results_cfs_ders}

\begin{table}[b]
\centering
\caption[]{ANOVA results for CFS--R and DERS--SF total scores. Holm-corrected $p$-values reported.}
\label{tab:anova_results_cfs_ders}
\small
\resizebox{\columnwidth}{!}{%
\begin{tabular}{llrrrr}
\toprule
\textbf{Outcome Variable (DV)} &
\textbf{Source} &
\textbf{$F$} &
\textbf{$p_{\text{unc}}$} &
\textbf{$p_{\text{Holm}}$} &
\textbf{Significance} \\
\midrule
CFS--R Total
& Condition
& 0.23 & .636 & 1.000 & No \\
& \textbf{Phase (pre--post)}
& \textbf{6.64} & \textbf{.020} & \textbf{.039}
& \textbf{Yes (Holm-corrected)} \\
& Phase $\times$ Condition
& 0.48 & .497 & 1.000 & No \\
\addlinespace
DERS--SF Total
& Condition
& 1.24 & .281 & 1.000 & No \\
& Phase (pre--post)
& 3.55 & .077 & .077 & Trend (uncorrected) \\
& Phase $\times$ Condition
& 1.32 & .267 & 1.000 & No \\
\bottomrule
\end{tabular}
}
\end{table}

\begin{table}[b]
\centering
\caption[]{Between-condition comparison of change scores ($\Delta$ Post--Pre) for all CFS--R and DERS--SF subscales. Medium or larger effects (|Hedges $g$| $\ge .50$) are bolded. \textcolor{black}{Blue values} with arrows indicate improvement for that condition (↑ = improvement on CFS--R; ↓ = improvement on DERS--SF).}
\label{tab:delta_between_conditions}
\small
\resizebox{\columnwidth}{!}{%
\begin{tabular}{lrrrr}
\toprule
\textbf{Subscale} &
\textbf{$\Delta$Free-form} &
\textbf{$\Delta$Gross-guided} &
\textbf{Hedges $g$} &
\textbf{Cohen's $d$} \\
\midrule
CFS--R Abandonment      & 0.90 & \textcolor{blue}{2.22 ↑} & -0.456 & -0.477 \\
CFS--R ReCoping         & \textcolor{blue}{0.60 ↑} & 0.44  &  0.050 &  0.052 \\
CFS--R Meta-Coping      & 1.40 & \textcolor{blue}{1.67 ↑} & -0.106 & -0.111 \\
\midrule
DERS Strategies         & -0.40 & \textcolor{blue}{-0.56 ↓} &  0.107 &  0.112 \\
\textbf{DERS Nonaccept} & 0.20  & \textcolor{blue}{-1.78 ↓} & \textbf{0.726} & \textbf{0.760} \\
DERS Impulse            & -0.50 & \textcolor{blue}{-0.67 ↓} &  0.115 &  0.121 \\
\textbf{DERS Goals}     & 0.20  & \textcolor{blue}{-1.56 ↓} & \textbf{0.598} & \textbf{0.626} \\
\textbf{DERS Awareness} & \textcolor{blue}{-0.70 ↓} & 0.44  & \textbf{-0.537} & \textbf{-0.563} \\
DERS Clarity            & 0.00  & \textcolor{blue}{-0.56 ↓} &  0.374 &  0.392 \\
\bottomrule
\end{tabular}
}
\end{table}

\begin{table}[b]
\centering
\caption[]{Post-hoc Phase effects for the Coping Flexibility Scale--Revised (CFS--R) subscales with Holm correction. The ``unc'' column reports uncorrected $p$-values. Effects that remain significant after Holm correction are shown in \textbf{bold}.}
\label{tab:cfs_phase_posthoc}
\small
\resizebox{\columnwidth}{!}{%
\begin{tabular}{lrrrrc}
\toprule
\textbf{Subscale} &
\textbf{$F$} &
\textbf{$p_{\text{unc}}$} &
\textbf{$p_{\text{Holm}}$} &
\textbf{Significance} &
\textbf{$\eta^2_p$} \\
\midrule
Abandonment
& 5.768 & \textbf{.028} & .056
& Trend (n.s. after Holm) & .253 \\
Re-Coping
& 0.586 & .454 & 1.000
& No & .033 \\
Meta-Coping
& 7.647 & \textbf{.013} & \textbf{.039}
& \textbf{Yes (survives Holm)} & .310 \\
\bottomrule
\end{tabular}
}
\end{table}

\begin{figure*}[t]
  \centering
  \includegraphics[width = 0.8\linewidth]{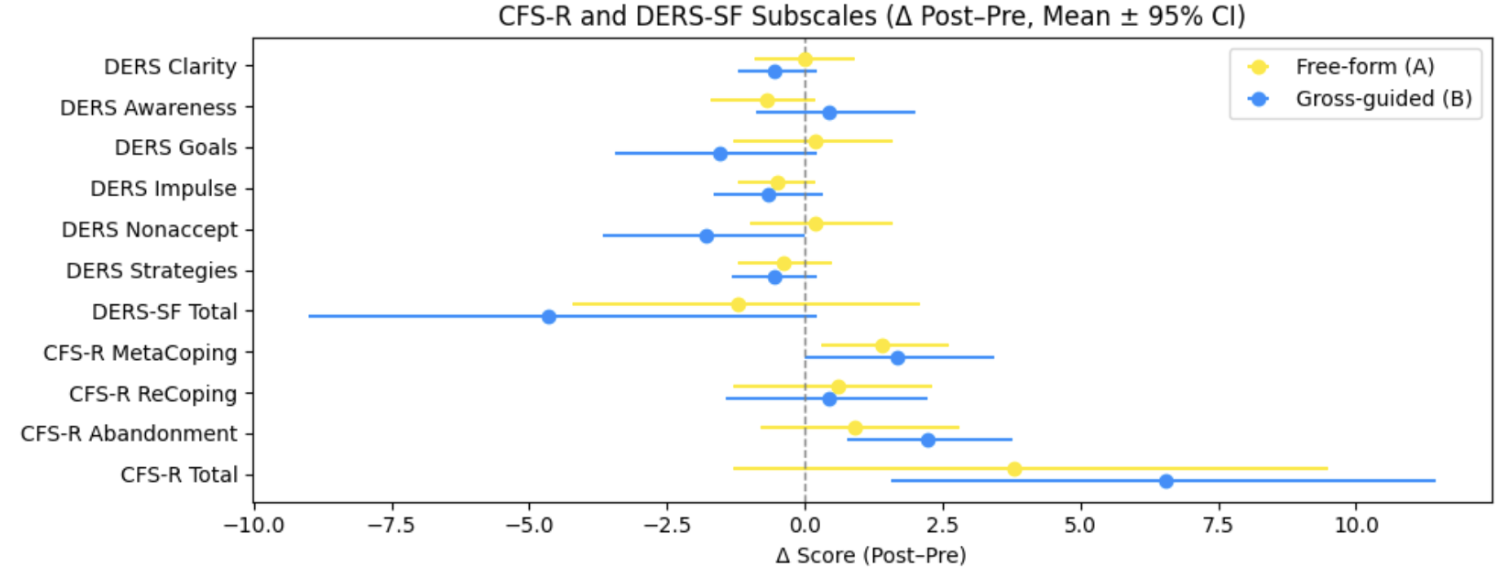}
  \caption[]{\textcolor{black}{Post–Pre $\Delta$ scores (bootstrapped 95\% CIs) for CFS--R and DERS--SF subscales under each condition (Gross-guided vs Free-form). Positive CFS--R $\Delta$ scores indicate improved coping flexibility, while negative DERS--SF $\Delta$ scores indicate reduced regulation difficulties. Both groups showed significant improvements in overall coping flexibility (RQ1). No between-condition differences reached statistical significance ($p > .12$). Gross-guided group showed medium-to-large reductions in \textit{Nonacceptance} and \textit{Goals} difficulty subscales, while the Free-form group showed medium-to-large reductions in \textit{Awareness} difficulty (RQ2).}}

  \label{fig:subscale_forest}
\end{figure*}

We examined overall intervention effects using mixed ANOVAs on \textcolor{black}{CFS--R and DERS--SF total scores (Table~\ref{tab:anova_results_cfs_ders}) and conducted exploratory analyses of subscale-level $\Delta$ (Post--Pre) change scores using Welch’s $t$-tests and Hedges’ $g$ (Table~\ref{tab:delta_between_conditions}; Fig.~\ref{fig:subscale_forest}). Because only the CFS--R Phase Effect outcomes were significant, we conducted Holm-corrected post-hoc Phase analyses for the CFS--R subscales (Table~\ref{tab:cfs_phase_posthoc}). We examine the subscale-level comparisons to explore the proposed mechanisms of change. This approach clarifies which specific components of coping flexibility are driving the observed improvements, informing more targeted hypothesis generation and future study design.}

\paragraph{Coping Flexibility (RQ1).}
As shown in Table~\ref{tab:anova_results_cfs_ders}, there was a \textbf{significant main effect of Phase for \textit{CFS--R Total}}, \textcolor{black}{($F(1,17)=6.64$, $p=.020$, $\eta^2_p=.28$), which remained significant after Holm correction ($p_{\text{Holm}}=.039$). This finding indicates reliable post-pre improvements in overall coping flexibility across both conditions (Phase Effect). At the subscale level, Table~\ref{tab:cfs_phase_posthoc} shows that \textbf{\textit{Abandonment} exhibited a significant Phase effect} ($F=5.77$, $p=.028$, $\eta^2_p=.25$), becoming marginal after correction ($p_{\text{Holm}}=.056$). \textbf{\textit{Meta-Coping} demonstrated the strongest Phase effect} ($F=7.65$, $p=.013$, $\eta^2_p=.31$), surviving correction ($p_{\text{Holm}}=.039$), indicating increased metacognitive monitoring and adaptive strategy adjustment. No Condition or Phase$\times$Condition interactions were significant for any CFS--R outcome (Table~\ref{tab:anova_results_cfs_ders}). Consistent with these null interaction effects, the $\Delta$-score contrasts in Table~\ref{tab:delta_between_conditions} were non-significant  (all $p>.12$). However, Fig.~\ref{fig:subscale_forest} shows medium-sized effect patterns favoring the Gross-guided group for \textit{CFS--R Total}, \textit{Abandonment}, and \textit{Meta-Coping}, suggesting that, on average, participants in the Gross-guided condition experienced greater improvement in coping flexibility, relative to participants in the Free Form condition.}

\paragraph{Emotion Regulation Difficulties (RQ2).}
\textit{DERS--SF Total} scores were non-significant (all $p>.26$) as in Table~\ref{tab:anova_results_cfs_ders}. Exploratory $\Delta$ contrasts (Table~\ref{tab:delta_between_conditions}) revealed several medium-sized, though non-significant effects: improvements favoring the \textbf{Gross-guided condition for \textit{Nonacceptance} ($g=0.73$) and \textit{Goals} ($g=0.60$)}, and a medium effect favoring the  \textbf{Free-form condition for \textit{Awareness} ($g=-0.54$)}. These complementary patterns, illustrated in Fig.~\ref{fig:subscale_forest}, suggest that Gross-guided prompts may reduce non-accepting responses and goal-disrupting regulatory barriers ($\downarrow$), whereas Free-form journaling may be more supportive for cultivating moment-to-moment emotional awareness ($\uparrow$). Although these group contrasts were non-significant in our pilot sample, these findings provide theoretically coherent hypotheses and insight for a future, fully-powered study.

\subsection{Weekly WhatIf-Planning Interaction and Readiness-for-Action (RQ1)}
\label{sec:results_readiness_for_action}

% Weekly WhatIf Review Items (\S\ref{sec:measures-WhatIfReview}) captured how participants translated intentions into concrete action (Fig.~\ref{fig:weekly_whatif}), and how strongly the system supported participants' formation of situation-specific intentions (via System Support Items; Fig.~\ref{fig:weekly_system_support}). Together with system logs (Table~\ref{tab:behavioral-engagement}), these indicators characterize participants’ readiness-for-action trajectories (R1) and how scaffolding shaped them (R2). 

% We analyzed based on \S\ref{sec:analysis-weekly}.}

\begin{figure*}[t]
\centering
\includegraphics[width=0.8\linewidth]{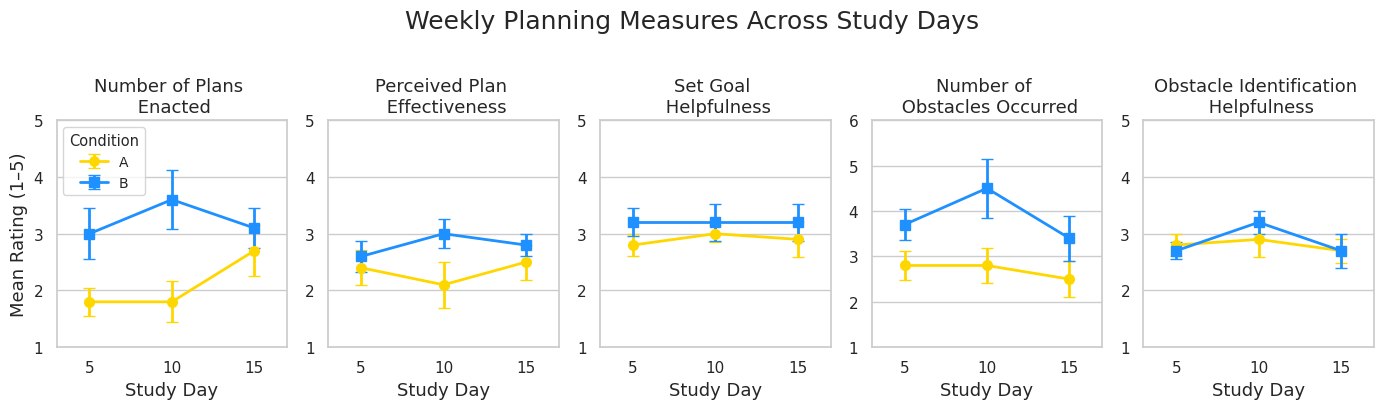}
\caption[]{
Weekly WhatIf-Planning Review outcomes (Days~5, 10, 15). 
Mean trajectories (with SEs) for \textit{Number of Plans Enacted}, \textit{Number of Identified Obstacles Occurred}, 
\textit{Perceived Plan Effectiveness}, \textit{Set Goal Helpfulness}, and \textit{Obstacle Identification Helpfulness}, \textit{Number of Obstacles Occurred}. 
These items index self-reported behavioral adoption and action-readiness (RQ1). 
No Condition$\times$Time interactions reached significance (all $p>.28$), 
but \textit{Plans Enacted} showed a significant main effect of Condition favoring the Gross-guided group.
}
\label{fig:weekly_whatif}
\end{figure*}

\begin{figure*}[t]
\centering
\includegraphics[width=0.8\linewidth]{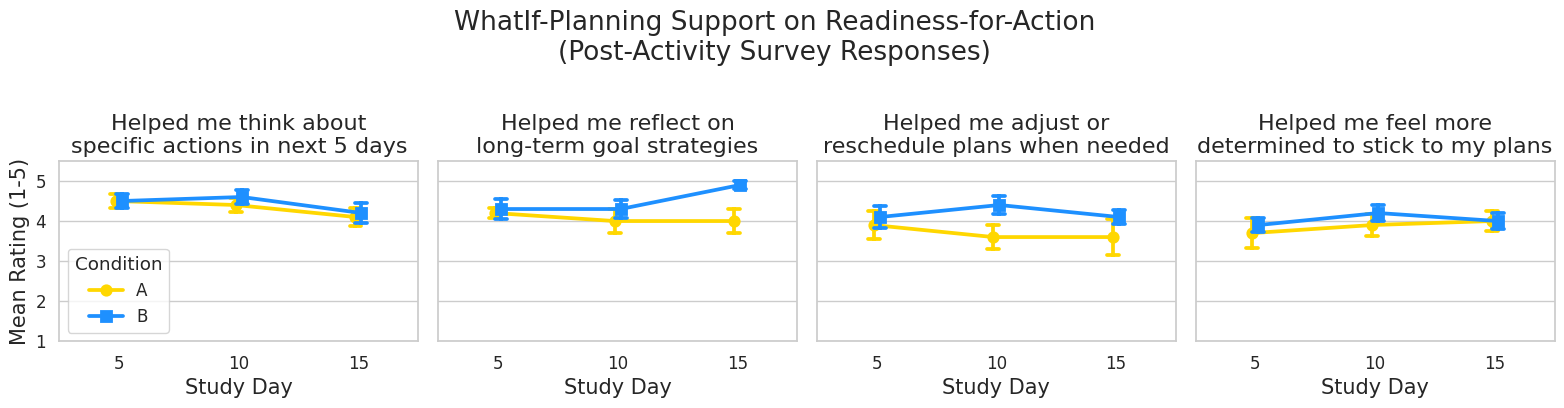}
\caption[]{
    Weekly system-support ratings for readiness-for-action (Days~5, 10, 15). Mean trajectories (with SEs) for obstacle identification, planning support, 
    short-term implementation intentions, and long-term strategy reflection. No significant Condition or Condition$\times$Time effects were detected;
    descriptively higher ratings in the Gross-guided condition parallel trends in metacognitive coping (RQ2).
  }
\label{fig:weekly_system_support}
\end{figure*}

\subsubsection{Weekly WhatIf Review Items.}
Piecewise linear mixed models on the Weekly WhatIf Review Items (\S\ref{sec:measures-WhatIfReview}) revealed a \textbf{significant main effect between conditions for Plans Enacted}, which is the main behavioral indicator for change. There were no significant Condition$\times$Time interactions across the other Weekly WhatIf Review Items (e.g., Set Goal Helpfulness; all $p>.28$). The piecewise mixed-effects model identified a significant main effect of Condition ($\beta = 1.20$, $SE = 0.57$, $z = 2.10$, $p = .036$), indicating that Gross-guided participants enacted approximately one additional plan per 5-day cycle relative to those in the Free-form condition. Furthermore, \textbf{the significant late-phase interaction} (seg2$\times$Condition: $\beta = -0.28$, $SE = 0.13$, $p = .034$) showed that although Free-form participants modestly increased enactment during Days~10--15, the Gross-guided group maintained higher behavioral follow-through throughout the 15-day study (Fig.~\ref{fig:weekly_whatif}). These findings demonstrate that while the two conditions had comparable perceptions of whether setting a goal was helpful, whether their identified obstacles occurred in real life, how helpful identifying obstacles was, and how effective their plans were, the participants in \textbf{Gross-guided condition had significantly more enacted plans in daily life}, suggesting that participants in this condition tended to have successful progress into the TTM Action stage (RQ1).

\vspace{-5pt}
\subsubsection{System Support Items (readiness-for-action support).}
Regarding how far participants perceived the platform to help them form concrete, situation-specific intentions (\S\ref{sec:measures-WhatIfReview}), participants' self-report ratings remained stable across Days~5, 10, and 15, with no significant Condition or Condition$\times$Time effects (Fig.~\ref{fig:weekly_system_support}). Descriptively, the Gross-guided condition showed slightly higher mean ratings; however, \textbf{there was no statistically significant difference in perceived system support between the two conditions}.

\begin{figure}[t]
  \centering
  \includegraphics[width = \linewidth]{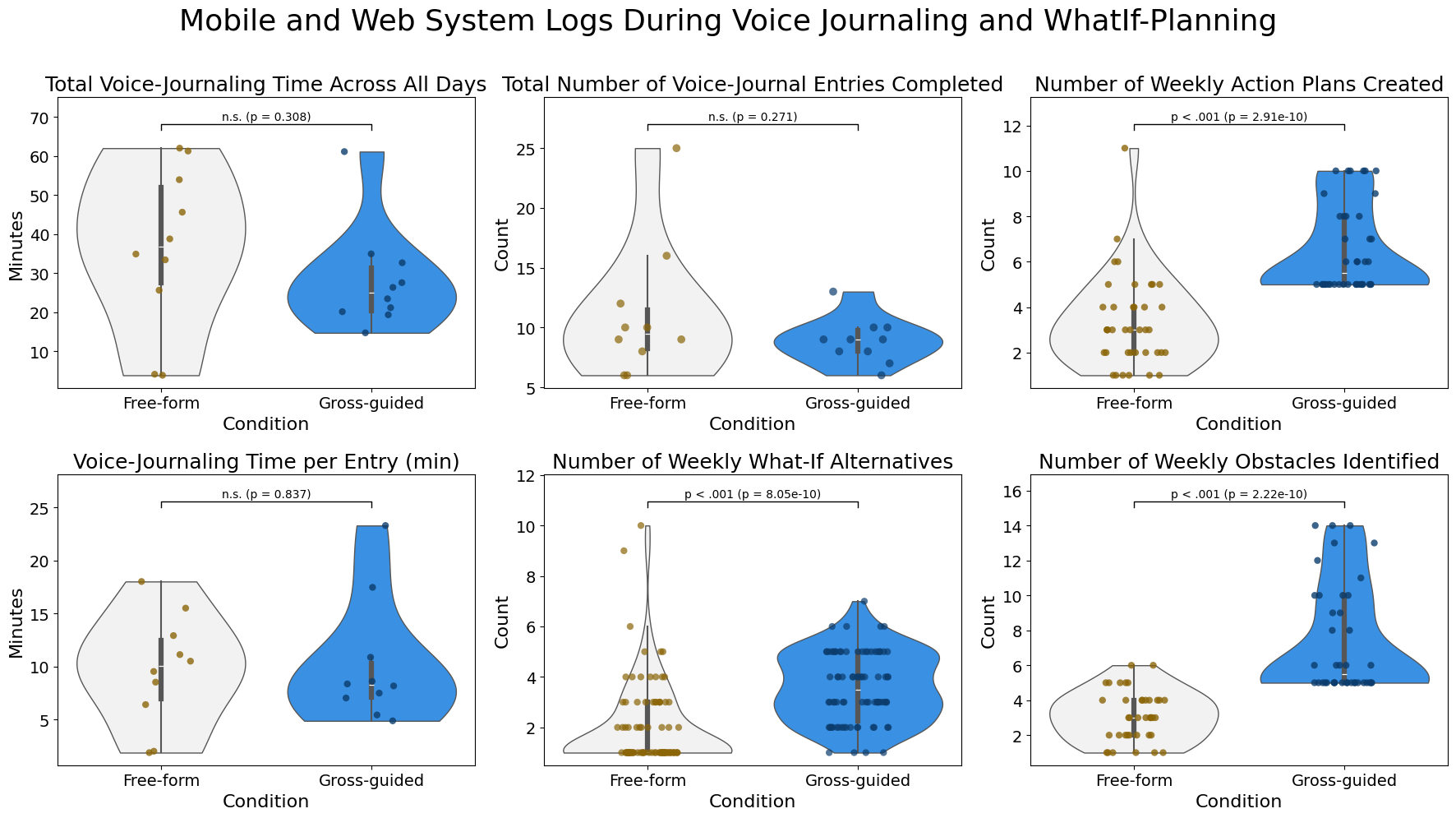}
  \caption[]{Daily journaling and WhatIf-Planning engagement by condition. Panels~1a--1c show daily voice-journaling behavior (total time, number of reflections, average duration). Panels~2a--2c show weekly planning outputs from WhatIf-Planning (counterfactuals, obstacles, action plans).}
  \label{fig:behavioral_engagement}
\end{figure}
  \vspace{-5pt}

\begin{table}[t]
\centering
\caption[]{Behavioral engagement with weekly WhatIf-Planning. Gross-guided group generated more alternatives,
identified more obstacles, and produced nearly twice as many action plans
as those in the Free-form condition, with all differences highly significant
($p < 10^{-9}$) and large effect sizes (Hedges $g = 0.99$--$1.71$).}
\label{tab:behavioral-engagement}
\small
\resizebox{\columnwidth}{!}{%
\begin{tabular}{lrrrrr}
\toprule
\textbf{System interaction}
& \textbf{Free-form}
& \textbf{Gross-guided}
& \textbf{$t$}
& \textbf{$p$}
& \textbf{Hedges $g$} \\
\midrule
What-If alternatives generated
& 2.03 & 3.59 & -6.53 & $\boldsymbol{8.05\times10^{-10}}$ & 0.99 \\

Obstacles identified
& 3.13 & 7.38 & -7.75 & $\boldsymbol{2.22\times10^{-10}}$ & 1.71 \\

Weekly action plans generated
& 3.33 & 6.53 & -7.25 & $\boldsymbol{2.91\times10^{-10}}$ & 1.62 \\
\bottomrule
\end{tabular}}
\end{table}

\subsubsection{WhatIf-Planning Weekly Interaction System Log}
Planning-related system logs provided objective indicators of how deeply participants engaged with the WhatIf-Planning workflow (\S\ref{sec:measures-weeklyInteraction}). As shown in Panels~2a--2c of Fig.~\ref{fig:behavioral_engagement} and Table~\ref{tab:behavioral-engagement}, Welch’s independent-samples $t$-tests revealed large and systematic differences favoring the Gross-guided condition. \textbf{Gross-guided} participants generated \textbf{significantly more What-If alternatives} ($M=3.59$ vs.\ $2.03$; $p=8.05\times10^{-10}$, $g=0.99$), \textbf{identified significantly more obstacles} ($M=7.38$ vs.\ $3.13$; $p=2.22\times10^{-10}$, $g=1.71$), and \textbf{produced nearly twice as many weekly action plans} ($M=6.53$ vs.\ $3.33$; $p=2.91\times10^{-10}$, $g=1.62$) than Free-form participants.

\subsection{Daily Voice Journaling Engagement (RQ3)}
\label{sec:results_journaling_engagement}

System logs showed that participants in both conditions engaged similarly with daily voice journaling (Fig.~\ref{fig:behavioral_engagement}, Table~\ref{tab:behavioral-engagement}, Panels~1a–1c). Free-form participants spent somewhat more total time journaling ($M=36.34$ vs.\ $28.13$~min), produced slightly more entries ($M=11.10$ vs.\ $8.90$), and recorded slightly shorter entries on average. However, \textbf{none of the differences in total time, time per session, or number of journaling sessions met traditional thresholds of statistical significance} ($p=.308$–$.837$, $g\approx 0.09$–$0.50$). Overall, the journaling intervention dose (frequency and duration of entries) was comparable across conditions. \textit{These findings suggest that downstream psychological or planning outcomes were unlikely to be driven by differences in total reflective time.}

Although these differences did not meet standard thresholds of statistical significance, the two conditions revealed notable distinct patterns of engagement. Free-form participants tended to journal more frequently ($p=.051$), whereas Gross-guided participants completed fewer but increasingly longer sessions, though the Condition~$\times$~Day interaction was not significant ($p=.050$).  

\vspace{-5pt}
\subsection{Overall System Usability (SUS) (RQ3)}
Both conditions rated the system above the established SUS benchmark of $68$ 
for acceptable usability~\cite{lewis2018system}.
The overall SUS score was $74.8$, with Free-form Condition averaging $73.8$ 
and Gross-guided Condition averaging $76.0$. 
Subscales indicated strong \emph{learnability} (Free-form = $80.6$, Gross-guided = $83.9$) 
and acceptable \emph{usability} (Free-form = $71.3$, Gross-guided = $73.9$). No significant differences emerged between conditions, suggesting that \textit{the added structure in the Gross-guided condition did not introduce additional friction or diminished user experience}.

\section{Qualitative Results} 
We conducted semi-structured interviews on Day~15 to examine participants’ experiences with the Reflection-to-Action workflow and its components: daily voice journaling, WhatIf-Planning, and the Gross-guided condition (\S\ref{sec:system-overview}). Interviews lasted approximately 20–40 minutes. Our aims were to understand: (i) how daily voice journaling and weekly WhatIf-Planning supported coping flexibility and progression toward action, or whether participants encountered any barriers (RQ1); (ii) whether and how participants’ experiences and perceived system effectiveness differed between conditions (Free-form vs.~Gross-guided) (RQ2); and (iii) how participants perceived and engaged with voice-based journaling, and whether these experiences differed across conditions (RQ2, RQ3). Full thematic codes, extended quotes, and interview questions are provided in Appendix~\ref{appendix:interview-questions}.

\subsection{Progression Toward TTM Action (RQ1)}

\begin{table*}[htbp]
\caption[]{Qualitative themes reflecting participant-reported affordances and limitations across system activities by condition. \% indicate the proportion of participants within each condition who referenced each theme.}
\label{tab:across-system-activities-most}
\small
\setlength{\tabcolsep}{6pt}
\renewcommand{\arraystretch}{1.12}
\begin{tabular}{@{}p{0.24\textwidth}p{0.24\textwidth}p{0.24\textwidth}p{0.24\textwidth}@{}}

\toprule
\multicolumn{4}{c}{\textbf{Across System Activities}} \\
\midrule
\multicolumn{2}{c}{\textbf{Most Liked}} & \multicolumn{2}{c}{\textbf{Most Disliked}} \\
\cmidrule(lr){1-2}\cmidrule(l){3-4}
\textbf{Free-form group} & \textbf{Gross-guided group}  & \textbf{Free-form group} & \textbf{Gross-guided group}  \\
\addlinespace[2pt]
\textit{\mbox{8 categories} (100\% reported)} &
\textit{\mbox{8 categories} (100\% reported)} &
\textit{\mbox{4 categories} (50\% reported)} &
\textit{\mbox{6 categories} (90\% reported)} \\
\midrule

% --- Most Liked: Free-form
\begin{minipage}[t]{\linewidth}
\raggedright
\begin{itemize}[leftmargin=*, nosep]
  \item Metacognitive insight into thought–emotion patterns (60\%)
  \item Emotional clarity (50\%)
  \item Ease of use, accessibility (40\%)
  \item Goal clarification and Planning support(40\%)
  \item Obstacle identification (30\%)
  \item Sense of ownership over reflection and decisions (30\%)
  \item Verbal processing (30\%)
\end{itemize}
\end{minipage}
&
% --- Most Liked: Gross-guided
\begin{minipage}[t]{\linewidth}
\raggedright
\begin{itemize}[leftmargin=*, nosep]
  \item Metacognitive insight into thought–emotion patterns (50\%)
  \item Verbal processing (40\%)
  \item Goal / Planning Support (30\%)
  \item Obstacle identification (30\%)
  \item Sense of ownership over reflection and decisions (30\%)
  \item Emotional clarity (30\%)
  \item Commitment to actions (30\%)
  \item Noticeable progress (30\%)
\end{itemize}
\end{minipage}
&
% --- Most Disliked: Free-form
\begin{minipage}[t]{\linewidth}
\raggedright
\begin{itemize}[leftmargin=*, nosep]
    \item Difficulty generating new alternatives over time (40\%)
  \item Difficulty in verbally reflecting (40\%)
  \item Onboarding difficulty (20\%)
  \item External constraints  (20\%)
\end{itemize}
\end{minipage}
&
% --- Most Disliked: Gross-guided
\begin{minipage}[t]{\linewidth}
\raggedright
\begin{itemize}[leftmargin=*, nosep]
   \item Difficulty generating new alternatives over time (22\%)
  \item Difficulty finding appropriate time or place to record (22\%)
  \item Unapplicable questions (22\%)
  \item Difficulty enacting plans (22\%)
  \item External constraints  (22\%)
  \item Unapplicable questions (11\%)
\end{itemize}
\end{minipage}
\\
\bottomrule
\end{tabular}
\end{table*}

\begin{table*}[htbp]
\caption[]{
Qualitative themes reflecting participant-reported affordances and limitations of \textit{WhatIf-Planning} by condition.}
\label{tab:whatif-op-likes-dislikes}
\small
\setlength{\tabcolsep}{6pt}
\renewcommand{\arraystretch}{1.12}
\begin{tabular}{@{}p{0.24\textwidth}p{0.24\textwidth}p{0.24\textwidth}p{0.24\textwidth}@{}}
\toprule
\multicolumn{4}{c}{\textbf{WhatIf-Planning}} \\
\midrule
\multicolumn{2}{c}{\textbf{Likes}} & \multicolumn{2}{c}{\textbf{Dislikes}} \\
\cmidrule(lr){1-2} \cmidrule(l){3-4}
\textbf{Free-form group} & \textbf{Gross-guided group} &
\textbf{Free-form group} & \textbf{Gross-guided group} \\
\addlinespace[2pt]
\textit{\mbox{6 categories} (90\% reported)} &
\textit{\mbox{5 categories} (90\% reported)} &
\textit{\mbox{4 categories} (70\% reported)} &
\textit{\mbox{4 categories} (60\% reported)} \\
\midrule

% Likes: Free-form
\begin{minipage}[t]{\linewidth}
\raggedright
\begin{itemize}[leftmargin=*, nosep]
  \item Obstacle identification (78\%)
  \item "Backup" plans or alternative strategies (70\%)
  \item Goal / Planning Support (44\%)
  \item Commitment to actions (22\%)
  \item Metacognitive insight into thought–emotion patterns (22\%)
  \item Self-efficacy (11\%)
\end{itemize}
\end{minipage}
&
% Likes: Gross-guided
\begin{minipage}[t]{\linewidth}
\raggedright
\begin{itemize}[leftmargin=*, nosep]
  \item Metacognitive insight into thought–emotion patterns (66\%)
  \item Obstacle identification (33\%)
  \item "Backup" plans or alternative strategies (33\%)
  \item Goal clarification (33\%)
  \item Broader perspective-taking (11\%)
  % \item \parbox[t]{\linewidth}{\raggedright Insight/\allowbreak Metacognition (66\%)}
  % \item \parbox[t]{\linewidth}{\raggedright Obstacle identification (33\%)}
  % \item \parbox[t]{\linewidth}{\raggedright Backup plans and adherence (33\%)}
  % \item \parbox[t]{\linewidth}{\raggedright Clarifies goals/plans (33\%)}
  % \item \parbox[t]{\linewidth}{\raggedright Broader perspective (11\%)}
  % \vspace{1pt}
\end{itemize}
\end{minipage}
&

% Dislikes: Free-form
\begin{minipage}[t]{\linewidth}
\raggedright
\begin{itemize}[leftmargin=*, nosep]
  % Dislikes: Free-form
  \item Difficulty generating new alternatives over time (57\%)
  \item Difficulty enacting plans (29\%)
  \item Need reminders (14\%)
\end{itemize}
\end{minipage}

&
% Dislikes: Gross-guided
\begin{minipage}[t]{\linewidth}
\raggedright
\begin{itemize}[leftmargin=*, nosep]
  \item External constraints  (33\%)
  \item Need reminders (10\%)
  \item Difficulty enacting plans (33\%)
  \item Delayed timing(17\%)
\end{itemize}
\end{minipage}
\\
\bottomrule
\end{tabular}
\end{table*}

\begin{table*}[htbp]
\caption{Qualitative themes reflecting participant-reported affordances and limitations of \textit{Voice Journaling} by condition.}
\label{tab:voice-journaling-likes-dislikes}
\small
\setlength{\tabcolsep}{6pt}
\renewcommand{\arraystretch}{1.12}
\begin{tabular}{@{}p{0.24\textwidth}p{0.24\textwidth}p{0.24\textwidth}p{0.24\textwidth}@{}}
\toprule
\multicolumn{4}{c}{\textbf{Voice Journaling}} \\
\midrule
\multicolumn{2}{c}{\textbf{Likes}} & \multicolumn{2}{c}{\textbf{Dislikes}} \\
\cmidrule(lr){1-2} \cmidrule(l){3-4}
\textbf{Free-form group} & \textbf{Gross-guided group}  & \textbf{Free-form group} & \textbf{Gross-guided group}  \\
\addlinespace[2pt]
\textit{\mbox{6 categories} (90\% reported)} &
\textit{\mbox{5 categories} (100\% reported)} &
\textit{\mbox{6 categories} (80\% reported)} &
\textit{\mbox{2 categories} (60\% reported)} \\
\midrule

% Likes: Free-form
\begin{minipage}[t]{\linewidth}
\raggedright
\begin{itemize}[leftmargin=*, nosep]
  \item Convenient, fast (56\%)
  \item"Stream-of-consciousness" (56\%)
  \item "Like talking to a friend" (44\%)
  \item Speech-to-text (22\%)
  \item Goal reminders (11\%)
  \item Novelty (11\%)

\end{itemize}
\end{minipage}
&
% Likes: Gross-guided
\begin{minipage}[t]{\linewidth}
\raggedright
\begin{itemize}[leftmargin=*, nosep]
  \item Convenient, accessible (70\%)
  \item Emotional clarity (40\%)
  \item Speech transcription (30\%)
  \item "Like talking to a friend" (20\%)
  \item"Stream-of-consciousness" (20\%)
  \vspace{1pt}
\end{itemize}
\end{minipage}
&
% Dislikes: Free-form
\begin{minipage}[t]{\linewidth}
\raggedright
\begin{itemize}[leftmargin=*, nosep]
  \item Desire more personalization (50\%)
  \item Preference for writing (25\%)
  \item Effort to read transcripts (25\%)
  \item Difficulty finding appropriate time or place to record (13\%)
  \item Unapplicable questions (13\%)
  \item External constraints  (13\%)
\end{itemize}
\end{minipage}
&
% Dislikes: Gross-guided
\begin{minipage}[t]{\linewidth}
\raggedright
\begin{itemize}[leftmargin=*, nosep]
  \item Difficulty finding appropriate time or place to record (100\%)
  \item External constraints  (17\%)
  % \item Finding place to record (100\%)
  % \item External factors (17\%)
  % \vspace{1pt}
\end{itemize}
\end{minipage}
\\
\bottomrule
\end{tabular}
\end{table*}

Participants across both conditions reported that WhatIf-Planning helped clarify goals, surface internal and external obstacles, and generate concrete, actionable strategies (Table~\ref{tab:across-system-activities-most}), indicating progression toward Action within the TTM.

Participants reported improved goal clarity (44\%), stronger awareness of alternative strategies (33\%), and greater plan-adherence (22\%). As one participant described:

\begin{quote}
\emph{``After writing my What-if alternatives, I remembered my plan in daily life. It felt easier to follow through.''}
\end{quote}

In Free-form condition, 78\% of participants reported that identifying recurring obstacles was helpful. One participant noted:

\begin{quote}
\emph{``Writing about what went wrong helped me realize it's always the same trigger.''}
\end{quote}

Participants in the Gross-guided condition described deeper metacognitive awareness. Two-thirds of participants in this condition (66\%) referenced benefits related to perspective-taking or reinterpreting their actions. For example:

\begin{quote}
\emph{``It forced me to stop blaming myself and think about other ways I could’ve handled it.''}
\end{quote}

One-third of Gross-guided participants (33\%) noted that structured prompts helped generate effective implementation strategies:

\begin{quote}
\emph{``Breaking it down by what I could control gave me something concrete to work with.''}
\end{quote}

Alongside these benefits, participants across both conditions described challenges that hindered consistent progress toward action. These included limited motivation or energy after work:  \emph{“Sometimes I just didn’t have the energy to reflect after work, even though I knew it might help.”}. Others noted difficulty remembering to engage with the system during busy workdays: \emph{“The hardest part was actually remembering to do it in the middle of my busy day.”}

A few also reported environmental and privacy-related constraints associated with voice journaling: 
\emph{“Finding a quiet space to talk was a challenge. I live with family, so I felt self-conscious.”}. Some expressed discomfort with being fully candid, knowing that recordings were stored: \emph{“I found it difficult to be honest at times, knowing someone might read my recordings.”}.

\vspace{-3pt}
\subsection{Gross-Guided vs.~Free-Form (RQ2)}

Although participants in both conditions found WhatIf-Planning useful, they described meaningful differences in how reflection unfolded. Participants in the Gross-guided condition reported more systematic reasoning and more emotionally regulated reinterpretations, often attributing these differences to the stepwise prompts aligned with Gross’s ER strategies.

Participants in the Gross-guided condition described increased awareness of unproductive fixation or rumination:

\begin{quote}
\emph{``Thinking through the steps made me realize I was fixating on one moment.''}
\end{quote}

They also reported heightened metacognitive awareness:

\begin{quote}
\emph{``I started recognizing patterns in how I react. The structure helped surface that.''}
\end{quote}

In contrast, participants in the Free-form condition more frequently reported moments when reflection felt stagnant:
\begin{quote}
\emph{“I often repeated myself \dots I wasn’t moving forward.”}.
\end{quote}

Interviews also revealed distinct affordance patterns: free-form group primarily used voice journaling to clarify goals and identify obstacles, whereas Gross-guided group more often emphasized its role in supporting targeted cognitive reframing and strategy generation. 

\vspace{-3pt}
\subsection{Perception on Voice Journaling (RQ3)}

Participants in both conditions appreciated the convenience and expressiveness of voice journaling compared to traditional text-based methods. All participants in the Gross-guided condition (100\%) and most participants in the Free-form condition (90\%) reported positive experiences with voice journaling. Across conditions, participants described the experience even when recording alone as \textit{“feeling like talking to a friend,”}  and noted that voice journaling often enabled \textit{“stream-of-consciousness.”}

Gross-guided participants also attributed increased emotional clarity (40\%), as illustrated by:

\begin{quote}
\emph{``Saying out loud helped me realize what I was feeling.''}
\end{quote}

However, voice-based interaction also had limitations. Across conditions, participants reported difficulty finding private places to record (60\%). Others noted challenges remembering to journal or reflecting long after an incident occurred:
\begin{quote}
\emph{``Sometimes I forgot, or didn’t know what to say without a prompt.''} and \emph{``I liked the prompt structure, but sometimes it felt delayed—like I needed to reflect earlier.''}
\end{quote}

\section{Discussion} 
In this section, we synthesize quantitative and qualitative findings to interpret how the Reflection-to-Action system supports adaptive coping and readiness for action \textbf{(RQ1)}, how Gross-guided reflection shapes emotion regulation and progression toward action \textbf{(RQ2)}, and how participants engaged with voice journaling in terms of usability and experience \textbf{(RQ3)}.

\vspace{-10pt}
\subsection{Reflection-to-Action System support Readiness for Action and Coping (RQ1)}

Across conditions, participants showed measurable gains in adaptive coping over the 15-day study. Overall, coping flexibility improved significantly (pre–post increase in \textit{CFS–R Total}; significant Phase effect). Post hoc analyses revealed significant gains in \textit{Meta Coping} ($F = 7.65$, $p_{\mathrm{Holm}} = .039$), indicating that participants became more deliberate in monitoring and adjusting their emotional responses over the course of the intervention. Improvements in \textit{Abandonment}—the ability to disengage from ineffective coping strategies—further suggest increased willingness to consider alternative responses. Together, these patterns align with coping flexibility theory~\cite{Kato2020frontiers, kato2020cfsr}, which posits that flexible selection and revision of coping strategies buffers against stress and supports adaptive outcomes. These indicate that the intervention can meaningfully support regulatory change, even within a 15-day window.

Weekly readiness measures clarify how these psychological gains translated into behavior. While subjective weekly ratings of plan effectiveness, goal-setting support, and obstacle identification were comparable across conditions, behavioral enactment differed substantially. Participants in the Gross-guided condition enacted significantly more plans than those in the Free-form condition—approximately three additional enacted plans per participant over the study period. System logs further illuminate this difference: the Gross-guided condition produced substantially richer planning output, including more What–If alternatives ($g = 0.99$), more obstacles identified ($g = 1.71$), and nearly twice as many weekly action plans ($g = 1.62$). A trend toward higher perceived plan effectiveness in the Gross-guided group (Fig.~\ref{fig:weekly_whatif}) suggests that participants also viewed these plans as more helpful for daily behavior. Together, these findings indicate that Gross-guided participants not only generated more structured plans but also followed through on them more consistently, reflecting higher behavioral implementation and readiness for action.

\textbf{Design implications.} Collectively, these results provide preliminary evidence that the WhatIf-Planning module can support the \textit{Preparation} stage of the Transtheoretical Model. Moreover, they suggest that embedding emotion regulation structure—via the Gross ER Process Model—into reflective interventions can increase plan enactment. Participants who reported recurring difficulties and unmet wishes began translating reflective insights into concrete, self-directed actions within a short time frame.

\vspace{-10pt}
\subsection{Reflection Structure Shapes Emotional Processing (RQ2)}

Although both conditions supported adaptive coping, they did so through distinct emotion-regulation pathways. Exploratory analyses of DERS–SF subscales showed that the Gross-guided group exhibited medium-sized improvements in \textit{Nonacceptance} ($g = 0.73$), reflecting reduced self-critical responses to emotional experiences, and in \textit{Goals} ($g = 0.60$), indicating greater capacity to remain goal-focused while distressed. In contrast, the Free-form condition showed a medium-sized improvement in \textit{Awareness} ($g = -0.54$), suggesting enhanced moment-to-moment emotional attunement. These differences indicate that structured prompts encouraged appraisal and goal alignment, whereas open-ended reflection fostered emotional noticing and experiential awareness.

Qualitative accounts reinforced this pattern. Participants in the Gross-guided condition described breaking down situations, reframing interpretations, and focusing on controllable elements—behaviors consistent with CFS–R metacognitive coping (i.e., evaluating responses and shifting away from ineffective strategies). Participants in the Free-form condition emphasized spontaneous expression, emotional release, and “hearing themselves think,” aligning with gains in emotional awareness and experiential processing.

\textbf{Design implications.} Taken together, these findings indicate that reflection structure meaningfully shapes emotion-regulation processes. Guided scaffolds channel reflection toward metacognitive evaluation and strategy-oriented reasoning, while Free-form reflection supports experiential processing and emotional exploration. This suggests that reflective technologies should align scaffold type with users’ regulatory needs—strengthening goal-directed regulation through structured prompts or enhancing emotional awareness through open-ended journaling. Future systems may benefit from adaptively shifting between these modes as users’ needs evolve.

\vspace{-8pt}
\subsection{Reflection Quality Over Quantity (RQ3)}

Participants across both conditions spent comparable total time journaling and rated the system as highly usable (overall SUS = 74.8). Thus, the additional structure in the Gross-guided condition neither required more reflection time nor imposed greater burden, despite yielding greater outcomes.

However, engagement patterns diverged. Free-form participants journaled more frequently and produced more entries overall, though these entries tended to be shorter (Table~\ref{tab:behavioral-engagement}) and were often described as “stream-of-consciousness” reflections (Table~\ref{tab:voice-journaling-likes-dislikes}). In contrast, Gross-guided participants made fewer but progressively longer entries over time, reflected in a significant Condition × Day interaction ($p = .050$). Their reflections also yielded significantly more weekly WhatIf-Planning outputs. Together, these patterns suggest that structured prompts can support deeper unpacking of situations and more sustained metacognitive reasoning per session, whereas Free-form journaling supports frequent emotional expression.

This clarifies the relative advantage of the Free-form condition on the \textit{DERS–SF Awareness} subscale: frequent open narration may enhance sensitivity to emotional cues, while Gross-guided prompts direct attention toward appraisal and goal alignment. These complementary affordances suggest that Free-form reflection may support emotional attunement, whereas structured reflection may better support metacognition and action-oriented insight.

\textbf{Design implications.} Overall, these results suggest that the \emph{quality} of reflection may matter more than its quantity. While many journaling tools emphasize reminders and frequency, our findings indicate that guided, high-quality reflection can yield more substantive regulatory gains without increasing reflection time.

\vspace{-4pt}
\section{Limitation and Future Work}

\paragraph{\textbf{From Feasibility to Full-Scale Impact.}} The main limitation of this pilot intervention is its small sample size. Instead, the study yielded rich longitudinal data ($20$ focal goals, $147$ daily voice recordings, and $192$ counterfactual action plans generated through WhatIf-Planning), as participants engaged in daily reflection over 15 days.

Although the study duration was short, we observed consistent patterns across primary outcomes, including coping flexibility and emotion regulation. Still, the limited sample size ($N=20$) and study duration constrained statistical power for subscale-level analyses. To address this, we examined effect sizes and triangulated quantitative measures with weekly review reports, system logs, and qualitative interviews to characterize early mechanisms of feasibility. In addition, our pilot study excluded individuals with self-reported clinical diagnoses and included only participants whose scenarios were feasible to address within a 15-day window. Future work with larger and more diverse samples (e.g., individuals with clinical diagnoses or those prone to maladaptive rumination) can examine whether the system can support broader mental health outcomes, such as depression or anxiety.

\paragraph{\textbf{Longer Deployments Are Needed to Assess Sustained Change.}}
Although measurable changes emerged within the 15-day intervention, this duration is insufficient to assess long-term regulatory trajectories or habit formation. Future studies should examine longer deployments to evaluate whether participants continue refining plans, adapting strategies to contextual changes, and sustaining coping gains. Richer longitudinal data (e.g., more ecological momentary assessments and passive sensing beyond self-report) could further clarify how reflection and planning translate into real-world behavior over time. Notably, two participants requested continued access to the system even without compensation, suggesting interest in longer-term use and high usability of our system.

% \paragraph{\textbf{Voice Journaling Was Effective, but Modality Comparisons Are Needed.}}
% Although voice journaling was not compared against a text-based baseline, participants consistently reported positive experiences with the voice modality, often describing increased self-disclosure and emotional clarity despite initial hesitation. Future work should directly compare voice-based and text-based journaling (e.g., daily written diaries) to determine whether similar reflective and regulatory benefits emerge across modalities. Additionally, analysis of acoustic features (e.g., pitch variability, pauses, or jitter) could provide further insight into emotional intensity and coping trajectories when paired with momentary assessments. Controlled studies are needed to validate these modality-specific effects.

\paragraph{\textbf{Need for Controlled Studies on Voice Journaling Effects.}}
The majority of participants reported positive experiences with the voice modality by the end of the study, often noting a shift from initial hesitation toward greater perceived convenience and emotional clarity. Participants also described increased convenience and self-disclosure when reflecting aloud. Further analysis of acoustic features (e.g., pitch variability, pauses, or jitter) could help characterize how voice-based journaling captures emotional intensity and coping dynamics. While our ad hoc analyses indicated that participants had prior experience with text-based journaling but none with voice journaling, future work should directly compare voice-based and text-based journaling to confirm whether similar reflective and regulatory benefits emerge across modalities.

% \vspace{-5pt}
\paragraph{\textbf{Intervention Arms Should Adapt to Evolving User Needs.}}
Moreover, while our design grounded prompts in stage-based theories of change (TTM and Gross’s Process Model), the conditions remained static once assigned. Participants may also be at different stages (e.g., contemplation vs. preparation), as their regulatory needs evolve over time. Future systems should explore adaptive, stage-sensitive scaffolding that dynamically calibrates reflection depth, timing, and modality. Additionally, while our system emphasized individual reflection, behavior change can be socially situated. Future work can investigate how journaling and planning might be embedded within relational contexts where self-driven change might be limited, and consider how scaffolds can be tailored to different scenarios or stages of change over longer deployments, while preserving privacy and user agency.

\paragraph{\textbf{Could AI Support Action Planning?}}
Interviews revealed a need for additional guidance during WhatIf-Planning, particularly in the Free-form condition, where participants reported difficulty generating new alternatives or plans over time. In response, we are planning a larger-scale study incorporating an AI-augmented condition to examine whether AI support can help users expand alternatives, obstacles, and action plans beyond self-generated input, potentially strengthening behavioral and emotional outcomes. A key challenge will be designing such support to enhance reflection and planning \textit{without undermining human agency}, especially for systems intended to foster mindful, self-directed behavior change.

\vspace{-4pt}
\section{Conclusion}
We present a \textbf{Reflection-to-Action system} (\S\ref{sec:system-overview}) that captures daily regrets and wishes through voice journaling and translates these reflections into coping strategies through the WhatIf-Planning module, which integrates counterfactual “what-if’’ review of captured voice-journal entries with “if–then’’ action planning. Grounded in the Transtheoretical Model (\S\ref{sec:theory-TTM}), we designed \textcolor{black}{one of the first systems to explicitly scaffold the} \textit{Preparation} stage and help individuals in the \textit{Contemplation} stage begin progressing toward \textit{Action}.  To our knowledge, this is also the first system to empirically operationalize Gross’s Emotion Regulation (ER) Process Model (\S\ref{sec:theory-Gross}) as a stepwise scaffold for reinterpretation, self-regulation across multiple leverage points, and everyday goal-directed action.

Across a 15-day in-the-wild study using triangulated system logs, self-reports, and interviews, we found multiple significant evidence that our Reflection-to-Action system supported adaptive coping and early progression toward action. Coping flexibility improved significantly across both conditions \textbf{(RQ1)}. While exploratory, the Gross-guided condition showed medium-to-large effects on several emotion-regulation subscales (DERS-SF), alongside significantly greater readiness for action, plan enactment, and WhatIf-Planning output \textbf{(RQ2)}. Participants in the Gross-guided condition enacted significantly more plans each week and generated more counterfactual alternatives, identified more obstacles, and authored nearly twice as many action plans compared to those in the Free-form condition \textbf{(RQ1, RQ2)}. Usability and interview data further indicated that these benefits were achieved without increased burden or reduced user preference for Gross-guided prompts \textbf{(RQ3)}. 

Taken together, this work makes four contributions:
(1) a unified \textbf{Reflection-to-Action framework} with a \textbf{WhatIf-Planning} module that scaffolds the \textit{Preparation stage} of the Transtheoretical Model, enabling individuals to capture momentary reflective insights and translate them into actionable plans that support early behavior change;
(2) the first empirical investigation of \textbf{Gross-guided} reflection as a structured scaffold for emotion regulation and coping flexibility;
(3) one of the first in-the-wild studies of \textbf{voice-based journaling}, characterizing usability and engagement patterns over time; and
(4) mixed-methods evidence from a 15-day in-the-wild study demonstrating that \textit{coping flexibility} can be improved through mindful reflection across conditions, and that \textbf{Gross-guided reflection substantially increases counterfactual generation, planning depth, and plan enactment}, informing the design of future behavior-change interventions.
By turning \textbf{reflection into action for self-driven change}, this work illustrates how interactive systems can move beyond passive recording to actively support insight formation, intention setting, and sustained self-directed growth in everyday life.

\begin{acks}
    This work was supported in part by the National Institute of Mental Health of the National Institutes of Health under Institutional Training Grant 5T32 MH125815-04. The content of this article is solely the responsibility of the authors and does not necessarily represent the official views of the National Institutes of Health. Statistical support was provided by data science specialist Joshua Cetron, at IQSS DSS, Harvard University. We also thank Dr. Petr Slovak for his valuable feedback on this work.
\end{acks}

\bibliographystyle{ACM-Reference-Format}
\bibliography{submission}

\appendix
\newpage
\section*{Supplementary Materials}

This supplementary section provides additional methodological, system, and qualitative details that support the main paper. We include (1) a full description of the system pipeline and backend logs, (2) pre-survey design and eligibility rationale, (3) complete system prompts by condition, (4) semi-structured interview protocols, (5) thematic codes with representative participant quotes, and (6) full daily and weekly survey items. 

\section{Engaged System Prompts by Study Condition}~\label{appendix:prompts-full}
This section documents the full set of prompts participants engaged with during the study, organized by experimental condition. We include the complete prompts to support transparency and replicability, and to clarify how each condition operationalized self-reflection and planning processes.

\begin{itemize}
  \item \textbf{Free-form condition prompts.} Participants in the \textbf{Free-form condition} were prompted with open-ended journaling questions (Table~\ref{tab:prompts-free-form}). These prompts encouraged reflection without specific guidance on emotion regulation stages.

  \item \textbf{Gross-Guided condition prompts.} Participants in the \textbf{Gross-Guided condition} were prompted with structured questions (Table~\ref{tab:prompts-grossGuided}) designed to align with Gross’s five families of emotion regulation strategies: situation selection, situation modification, attention deployment, cognitive change, and response modulation.
\end{itemize}

\begin{table}[b]
\centering
\caption{Prompts for Free-form condition mapped to each reflection-to-action system component.}
\label{tab:prompts-free-form}
\small
\setlength{\tabcolsep}{6pt}
\renewcommand{\arraystretch}{1.25}
\begin{tabular}{@{}p{2.7cm}p{0.65\linewidth}@{}}
\toprule
\textbf{Activity} & \textbf{Prompt} \\
\midrule
\textbf{Voice Journaling} &
Please describe in detail, without naming anyone: \textbf{what} happened, \textbf{who} was involved, \textbf{when} and \textbf{where} it took place, \textbf{how you felt}, and \textbf{what you tried}. \\
\midrule
\textbf{What-If} &
What could you have done \textbf{differently}? \\
\midrule
\textbf{If--Then (Obstacle)} &
What is it within you—your \textbf{actions, reactions}, or \textbf{thoughts}—that most \textbf{hold you back} from fulfilling this wish? \\
\midrule
\textbf{If--Then (Planning)} &
List all the \textbf{alternative actions, reactions}, or \textbf{thoughts} you can do differently to \textbf{overcome} this obstacle. \\
\bottomrule
\end{tabular}
\end{table}

\begin{table*}[!htbp]
\centering
\caption{Mapping of prompts engaged in by Gross-Guided condition to each reflection-to-action system component}
\label{tab:prompts-grossGuided}
\small
\renewcommand{\arraystretch}{1.3}

\begin{tabular}{@{}p{2.2cm}p{0.16\linewidth}p{0.16\linewidth}p{0.16\linewidth}p{0.16\linewidth}p{0.16\linewidth}@{}}
\toprule
 & \textbf{Situation Selection} 
 & \textbf{Situation Modification} 
 & \textbf{Attention Deployment} 
 & \textbf{Cognitive Change} 
 & \textbf{Response Modulation} \\
\midrule

\textbf{Voice Journaling} 
& What is the situation you \textbf{engaged in or avoided}? 
& Did you do anything to \textbf{impact} how the situation unfolded, if any? 
& Can you remember what \textbf{caught your attention} or what you \textbf{focused} on in the situation? 
& How did you \textbf{interpret} the situation at the time? 
& Did you notice anything about \textbf{how you responded} emotionally or physically, or through your actions? \\
\midrule

\textbf{What-If} 
& How could you have \textbf{engaged/avoided} differently? 
& How could you have \textbf{impacted} differently? 
& How could you have \textbf{focused} on different elements of the situation? 
& How could you have \textbf{interpreted} differently? 
& How could you have \textbf{reacted} differently? \\
\midrule

\textbf{If--Then (Obstacles)}
& Could the way you are \textbf{selecting or avoiding} the situation be making things more challenging? 
& Could the way you’re \textbf{trying to change} the situation be making things more challenging? 
& Could what you are \textbf{focusing on} make the situation more challenging? 
& Could the way you’re \textbf{interpreting things} be making things more challenging? 
& Could the way you’re \textbf{reacting} to the situation be making things more challenging? \\
\midrule

\textbf{If--Then (Planning)}
& Is there a different way you could \textbf{select or avoid} the situation? 
& How can you \textbf{act differently to modify} the situation? 
& How can you \textbf{focus} differently? 
& How can you \textbf{interpret} the situation differently? 
& How can you \textbf{react differently} to the situation? \\
\bottomrule
\end{tabular}
\end{table*}

\section{Full System Pipeline and System Logs}~\label{Appendix:Full-System}

\begin{figure*}[tbp]
 \centering
 \includegraphics[width=\linewidth]{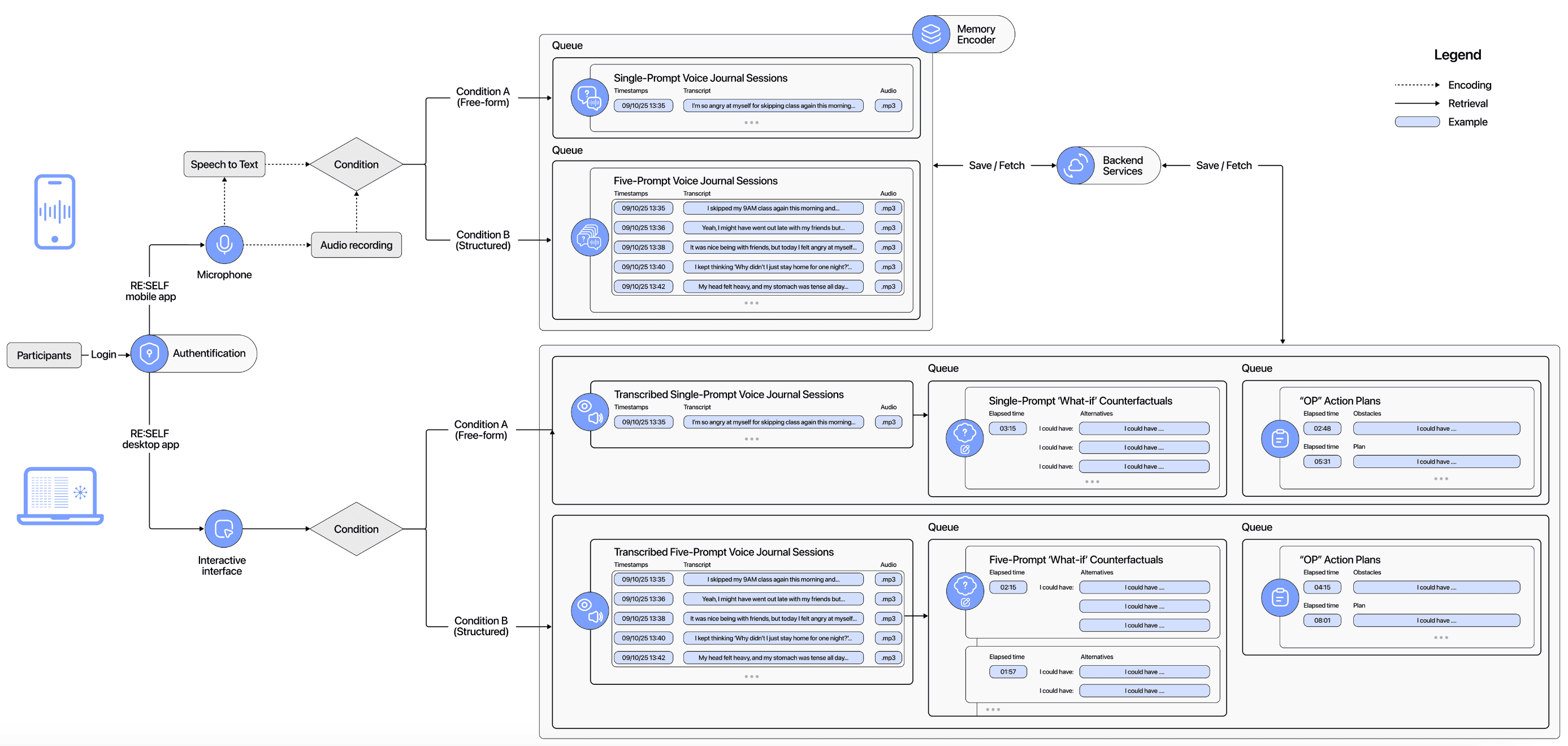}
 \caption{Overview of the system implementation. Participants journaled on mobile in either free-form or Gross-guided conditions. Saved entries were then accessible in the web app, where participants could review transcripts with audio, track progress on a calendar, and generate reflections with WhatIf-Planning module. The sequential modules include: journaling prompts, speech-to-text transcription, review of past journals, counterfactuals for each of the past journals, obstacle identification, and action planning.}
 \label{fig:SystemDiagram-full}
\end{figure*}

Figure~\ref{fig:SystemDiagram-full} presents the end-to-end system workflow, beginning with mobile voice journaling and continuing through web-based review, reflection, and action planning. Beyond illustrating the user-facing flow, the system also captures backend logs throughout the study, which were used to triangulate evidence related to user outcomes and system usability. This design ensured consistent delivery of study conditions while enabling fine-grained analysis of engagement and planning behaviors over time.

\section{Recruitment Survey Design and Insights}~\label{appendix:pre-survey}
This section provides the full pre-survey items and elaborates on the rationale for the focus group eligibility criteria. We also present insights from the pre-survey assessing participants’ perceptions of recurring regret and unmet wish scenarios. Participants were initially recruited through a Qualtrics survey distributed via mailing lists, and respondents were compensated \$1 for each entry.

Twenty participants who met the subject eligibility criteria in the screening survey and the focus group criteria in the pre-survey were recruited to participate in a 15-day study. The eligibility process was designed to ensure methodological rigor in participant selection by specifically recruiting individuals in the contemplation stage, as our intervention focuses on scaffolding the preparation stage of the Transtheoretical Model (TTM) to support progression from contemplation to action. This process also enabled us to identify common patterns in participants’ recurring regret themes to inform future intervention design.

\subsection{Screening Survey: Study Subject Eligibility}
To participate in the study, we set subject eligibility criteria to ensure their ability to safely and effectively engage with the study tasks. Participants were eligible if they were between 18 and 100 years old, fluent in English (reading, writing, speaking, and listening), and comfortable verbally reflecting on their thoughts and experiences, including emotionally difficult or recurring situations they might regret. They were also required to be comfortable using mobile apps and web-based tools and have regular access to a smartphone and reliable internet. \textcolor{black}{We also let participants know that they will be engaging in voice journaling, therefore need to be comfortable verbally reflecting and recording their journals.} Exclusion criteria included self-reported diagnoses of cognitive impairments or psychiatric conditions that would interfere with participation, current suicidal thoughts, or hearing and speech difficulties that would prevent verbal participation. Because the WhatIf-Planning and Gross-guided modules have not been clinically validated, we set these exclusion criteria to ensure participant safety and to allow us to examine feasibility and understand how these features operate before extending the work to clinical contexts. Finally, we ensured that there were no conflicts of interest with the principal investigators, and all study procedures were approved by the Institutional Review Board (IRB).

\subsection{Pre-survey: Focus Group Eligibility}~\label{appendix:pre-survey-items}
Pre-survey was used to recruit focus group in contemplation stage of TTM and to gain insights into the types of recurring regrets or unmet wishes individuals have as there is lack of publicly available data on recurring regret or unmet wishes scenarios with detailed ratings on importance, intensity, and perceived opportunity for change. Eligibility criteria were applied to ensure that regrets were both meaningful and actionable to make self-driven changes. 

A scenario was eligible if it: (1) occurred at least three times per week; (2) was rated moderate-to-high in emotional intensity ($\geq 3.5$ on a 5-point scale); (3) was perceived to have moderate-to-high opportunity for change ($\geq 3.0$); and (4) was not rated lowest in subjective importance. These criteria ensured that retained regrets were emotionally salient, personally meaningful, and realistically actionable. Items included:  

\begin{enumerate}
    \item \textbf{Domain Endorsement}
    \begin{itemize}
        \item ``In the following relationship domains, do you have any recurring wish or regret that is likely to happen again in the next two weeks and that you would like to try making changes in? (Select all that apply)''
        \begin{itemize}
            \item Romance (e.g., postponing plans, avoiding conversations, ignoring a partner’s message, toxic relationships)
            \item Friendship (e.g., cancelling plans, taking too long to reply, avoiding addressing an issue)
            \item Parent/Family (e.g., losing patience, ignoring a parent’s message, missing quality time, avoiding resolving a disagreement)
            \item Work/Academic Relationship (e.g., tension with coworkers or manager, communication issues with advisor, avoiding feedback)
        \end{itemize}

        \item ``In the following non-relationship domains, do you have any recurring wish or regret that is likely to happen again in the next two weeks and that you would like to try making changes in? (Select all that apply)''
        \begin{itemize}
            \item Academic/Work (tasks, deadlines, assignments)
            \item Health (exercise, sleep, diet)
            \item Finance (spending, bill payment, budgeting)
        \end{itemize}
    \end{itemize}

    \item \textbf{Importance}  
    ``How important is this part of your life?'' (1 = Not at all important, 5 = Extremely important)

    \item \textbf{Intensity}  
    ``How emotionally intense is this regret/wish?'' (1 = Extremely weak, 5 = Extremely intense)

    \item \textbf{Opportunity for Change}  
    ``How easy is it to change or modify this part of your life? Is this something you can act on vs.\ out of your control?''  
    (1 = Very hard to change, 5 = Very easy to change)
\end{enumerate}

\subsection{Insights from Pre-Survey Target Scenarios}~\label{Appendix:InsightsFromTargetScenarios}
\begin{figure}[b]
 \centering
 \includegraphics[width=\linewidth]{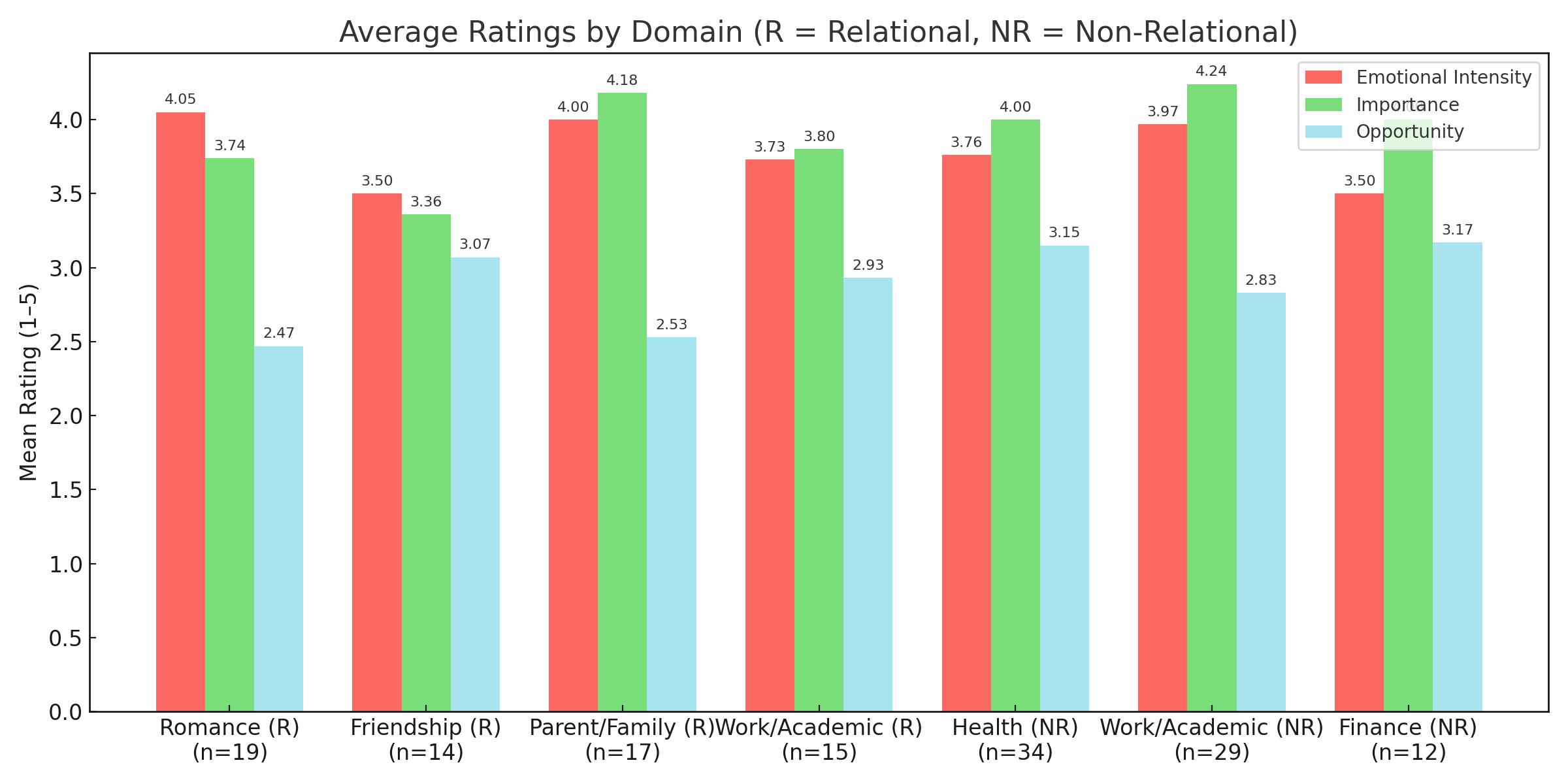}
 \caption{Mean ratings of Intensity, Importance, and Opportunity for Change (1–5) across relational (R) and non-relational (NR) domains, with sample sizes indicated under each domain. Participants ($n = 100$) can report at most 3 relational or non-relational recurring regret or wishes scenarios.}
 \label{fig:combined_domain_results_bar}
\end{figure}

In the pre-survey ($n=100$), participants described up to three recurring regrets or wishes across relational (R) and non-relational (NR) domains. Responses (rated 1-5) revealed variation in perceived intensity, importance, and opportunity for change (Figure~\ref{fig:combined_domain_results_bar}).

Within relational domains, romance scenarios were rated as both important (3.74) and moderately actionable (opportunity = 2.47). Parent/family scenarios stood out as the most important among all relational domains (importance = 4.18), with a moderate opportunity for change (2.53). Friendship scenarios showed mid-level importance (3.36) but somewhat higher opportunity (3.07), suggesting participants viewed them as areas with more potential for improvement. By contrast, work/academic relational scenarios were rated as less important (3.80) relative to family or romantic scenarios, despite being moderately intense.  

Among non-relational domains, finance displayed the strongest perceived opportunity for change (3.17), coupled with high importance (4.00). Health scenarios also scored high in importance (4.00) but showed relatively modest opportunity for change (3.15). Non-relational work/academic tasks were rated as both highly important (4.24) and moderately actionable (2.83).  

When comparing relational or non-relational wishes or regret domains, non-relational scenarios were rated somewhat higher in overall importance (M = 4.04) compared to relational domains (M = 3.77). However, relational domains included more scenarios that participants perceived as important and potentially actionable.  

Taken together, these results suggest that intervention may be most appropriate in domains where importance and opportunity converge at relatively high levels. Parent/family, finance, and romance stand out as particularly promising targets, as they combine strong personal relevance with perceived capacity for change. Friendship also shows potential, given its moderate importance but comparatively higher opportunity for improvement. In contrast, health and academic/work domains—though important—may present more structural barriers to change, making them less immediately suitable for short-term intervention. Notably, because only scenarios that recurred multiple times within a week were eligible for inclusion, fewer relational regrets or wishes met the final screening thresholds compared to non-relational domains.

\begin{figure}[b]
 \centering
 \includegraphics[width=\linewidth]{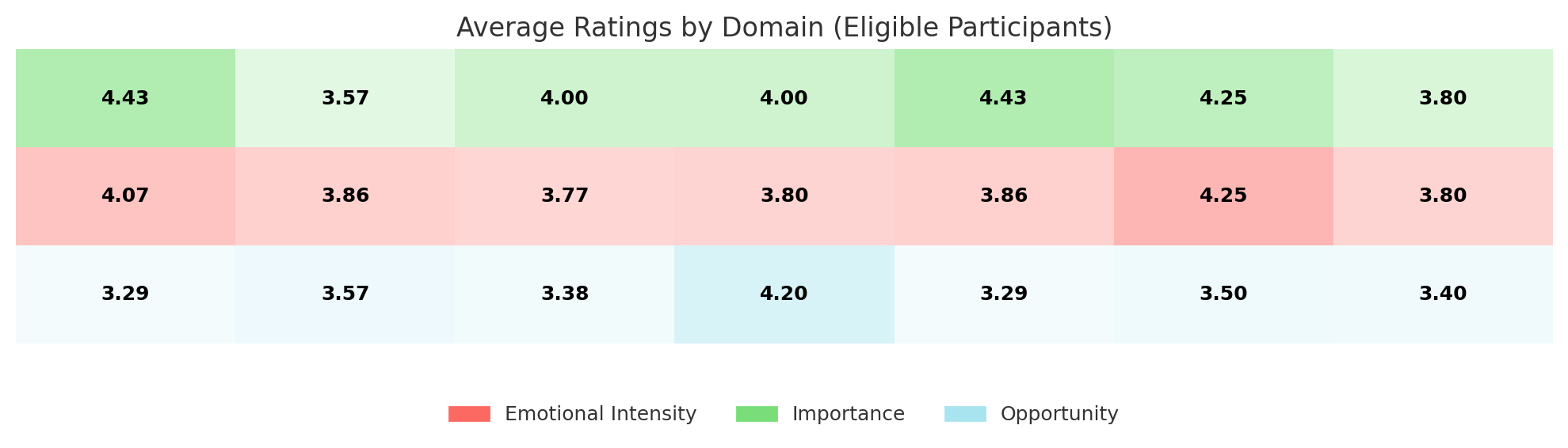}
 \caption{Heatmap of mean ratings (1–5) for \textit{Emotional Intensity}, \textit{Importance}, and \textit{Opportunity for Change}, separated by relational (R) and non-relational (NR) domains in the Focus group ($n=20$).}
 \label{fig:combined_domain_results_heatmap}
\end{figure}

\section{Daily and Weekly Surveys}

Full daily survey items (Table~\ref{tab:daily-survey}) and weekly survey items (Table~\ref{tab:weekly-survey}) assessed participants’ reflections, emotion regulation strategies, and usability. We only analyzed the most relevant constructs and measures and report the findings in this paper.

\begin{table*}[!htbp]
\centering
\caption[]{Daily survey questions on voice journaling (mobile). Constructs align with emotion regulation and usability frameworks.}
\label{tab:daily-survey} 
\small
\renewcommand{\arraystretch}{1.25}
\begin{tabular}{@{}p{0.8cm}p{0.65\textwidth}p{3.5cm}@{}}
\toprule
\textbf{Q\#} & \textbf{Question} & \textbf{Construct} \\
\midrule
Q1 & Today’s voice journaling helped me recognize behaviors or habits that got in the way of my goals. & Obstacle Identification \\
Q2 & Overall, how overwhelmed or stressed did you feel today? & Stress / Overwhelm \\
Q3 & Journaling today helped me feel clear (vs confused) about my feelings. & Emotional Clarity \\
Q4 & Did you try anything new today to manage your emotions or the situation? & Strategy Planning / Adoption \\
Q5 & Overall, how easy or difficult was it to do reflective activity today? & Usability – Mobile \\
\bottomrule
\end{tabular}
\end{table*}

\begin{table*}[!htbp]
\centering
\caption{Weekly survey questions on What-If and If--Then Planning (web). Constructs span obstacle identification, planning, and engagement.}
\label{tab:weekly-survey}
\small
\renewcommand{\arraystretch}{1.25}

\begin{tabular}{@{}p{0.8cm}p{0.65\textwidth}p{3.5cm}@{}}
\toprule
\textbf{Q\#} & \textbf{Question} & \textbf{Construct} \\
\midrule
Q1  & How many previously identified obstacles occurred in the last five days? & Obstacle Occurrence \\
Q2  & How helpful was identifying obstacles in addressing situations? & Obstacle Identification Effectiveness \\
Q3  & How many plans did you try in the last five days? & Behavior Adoption \\
Q4  & How effective were your plans in addressing situations? & Planning Effectiveness \\
Q5  & The If--Then (Action Planning) activity helped me consider rescheduling or adjusting plans. & Plan Adaptation \\
Q6  & The If--Then (Obstacle Identification) activity helped me identify obstacles that might interfere. & Obstacle Identification \\
Q7  & The If--Then (Action Planning) activity helped me feel determined to stick to plans. & Plan Adherence / Motivation \\
Q8  & The If--Then (Action Planning) activity helped me reflect on long-term strategies. & Long-Term Planning \\
Q9  & The If--Then (Action Planning) activity helped me plan specific actions for the next five days. & Short-Term Planning \\
Q10 & Ease of using the system during If--Then (Obstacle Identification). & Usability – Web \\
Q11 & Ease of using the system during If--Then (Action Planning). & Usability – Web \\
Q12 & Ease of using the system during the What-If activity. & Usability – Web \\
Q13 & Engagement with If--Then (Obstacle Identification). & Engagement \\
Q14 & Engagement with If--Then (Action Planning). & Engagement \\
Q15 & Engagement with the What-If activity. & Engagement \\
Q16 & What part of the interface could be improved? & Usability Improvements \\
\bottomrule
\end{tabular}
\end{table*}

\begin{table}[t]
\centering
\caption{Relational (R) and non-relational (NR) domains by study condition (Free-form vs Gross-guided) that met eligibility criteria, with participant counts. Both conditions included a balanced mix of R and NR domains.}
\label{tab:final_domain_types}
\begin{tabular}{lcc}
\hline
\textbf{Domain Type} & \textbf{Free-form} & \textbf{Gross-guided} \\
\hline
Non-Relational & 7 & 6 \\
Relational & 3 & 4 \\
\hline
\textbf{Total} & \textbf{10} & \textbf{10} \\
\hline
\end{tabular}
\end{table}

\section{Semi-Structured Exit Interview}~\label{Appendix:InterviewQuestions}

\textcolor{black}{We conducted semi-structured exit interviews (20--40 minutes) to better understand participants’ experiences with daily journaling, weekly planning, and the overall system. Questions were designed to elicit reflections on usability, perceived outcomes, and suggestions for future designs.}

\vspace{8pt}
\subsection{Interview Questions}
\label{appendix:interview-questions}
\paragraph{\textbf{Part 1: Daily Voice Journals (Mobile App)}}

\begin{enumerate}%%[label=Q\arabic*.]
    \item[Q1.] Can you describe the situations that prompted you to journal in the last 15 days, whether because you felt motivated, reminded, or triggered by something?
    \item[Q2.] Were there times you thought about journaling but decided to skip in the last 15 days? What made it easy or difficult to journal?
    \item[Q3.] What was it like to use your voice instead of writing for journaling? How was the experience different, if any?
    \item[Q4.] Did your experience with voice journaling vary depending on your environment (e.g., location, time of day, people around you)? If so, can you describe more?
\end{enumerate}

\paragraph{\textbf{Part 2: Weekly WhatIf-Planning (Web App)}}

\paragraph{\textit{What-If Review}}
\begin{enumerate}%%[resume, label=Q\arabic*.]
    \item[Q5.] What did you like or dislike about the ``What-If'' activity, coming up with alternative actions or reactions about your past journal?
    \item[Q6.] How easy or difficult was the experience of generating alternatives?
\end{enumerate}

\paragraph{\textit{If--Then Planning}}
\begin{enumerate}%%[resume, label=Q\arabic*.]
    \item[Q7.] What did you like or dislike about the If--Then planning activity (identifying obstacles and coming up with new strategies)?
    \item[Q8.] Have you ever tried to follow through on your weekly plans? If yes, how did it go? If not, what prevented you?
    \item[Q9.] Can you share a time when your plans didn’t work as expected? What did you do then?
\end{enumerate}

\paragraph{\textbf{Part 3: Overall Experience}}
\begin{enumerate}%%[resume, label=Q\arabic*.]
    \item[Q10.] Looking back at your Day 0 goal, has your perspective on recurring regrets or wishes changed?
    \item[Q11.] On a scale of 1--5, how close do you feel to attaining your goal or change?
    \item[Q12.] Which part of the system would you continue using without compensation, if any?
    \item[Q13.] After 15 days, how much control (1--5) do you feel you have over this topic?
    \item[Q14.] Do you believe you can continue making changes after this study? How confident are you (1--5)?
\end{enumerate}

\paragraph{\textbf{Part 4: Study Feedback \& Future Design}}
\begin{enumerate}%%[resume, label=Q\arabic*.]
    \item[Q15.] What do you think is the biggest challenge people face when trying to change regrets or wishes?
    \item[Q16.] If you could imagine your ideal tool for turning reflection into action, what would it look like or do?
    \item[Q17.] Looking back at the last 15 days, what’s one thing you liked most and one thing you disliked most?
    \item[Q18.] If AI were added to support reflection and behavior change, what role should it play? Are there parts of reflection you’d want to keep fully personal?
\end{enumerate}

\subsection{Thematic Codes with Representative Participant Quotes}~\label{Appendix:ThematicCodes}

\textcolor{black}{This appendix presents additional representative participant quotes corresponding to the qualitative themes reported in the Results section. Quotes are organized by research question to illustrate participants’ perceived outcomes (RQ1), experiences with structure versus freedom in reflection (RQ2), perceptions of the voice modality (RQ3), and broader challenges in sustaining self-reflection and behavior change. Together, these excerpts provide qualitative grounding for the reported themes and highlight variations in participant experiences across conditions.}

% \section{Appendix C. Thematic Codes and Raw Participant Quotes}

% -----------------
% RQ1
% -----------------

\subsubsection{\textbf{User Perceptions of Outcomes (RQ1)}}
Table~\ref{tab:quotes-rq1} presents selected quotes illustrating participants’ perceptions of varying outcome levels. For example, some participants struggled with repetition (“It was the same list that I just kept writing” — P75096-A), while others reported perceived growth (“It gave me a sense of direction instead of feeling lost” — P53782-B).

% -----------------
% RQ2
% -----------------
\subsubsection{\textbf{User Perception on Structure vs Freedom (RQ2)}}

Table~\ref{tab:quotes-rq2} presents selected quotes illustrating participants’ perceptions of structured versus free-form self-reflection. Several participants highlighted the benefits of free-form reflection for capturing unstructured thoughts and initiating a stream of consciousness (e.g., “I liked being able to just say whatever was on my mind” — P81956-A). In contrast, participants in the Gross-guided condition emphasized the value of structured guidance for supporting reframing and focused reflection (e.g., “The prompts reminded me to reframe” — P60981-B).

% -----------------
% RQ3
% -----------------
\subsubsection{\textbf{User Perception of Voice Modality (RQ3)}}

Table~\ref{tab:quotes-voice} presents selected quotes illustrating participants’ perceptions of voice journaling, including convenience, emotional clarity, and environmental constraints. Participants praised both convenience (“I could be on the bus and still record” — P37445-A) and clarity (“Speaking helped me realize what I was really feeling” — P81956-A). At the same time, environmental barriers were common (“It was hard to find a quiet, private space” — P26409-A). The free-form group emphasized spontaneity and personal discovery (P37445-A, P81956-A), while Gross-guided participants highlighted efficiency and structural supports (P37444-B, P60981-B).
% -----------------
% Background
% -----------------
\subsubsection{\textbf{Participant Challenges in Making Change with the System}}

Table~\ref{tab:quotes-challenges} presents selected quotes to contextualize broader challenges in sustaining self-reflection and behavior change beyond modality-specific perceptions. Participants described both motivational and environmental barriers. “Sometimes I just didn’t have the energy to reflect after work” (P37444-B) contrasted with concerns about honesty: “I found it difficult to be honest at times” (P26409-A). 

\subsubsection{\textbf{Synthesis Across Research Questions.}}
\textcolor{black}{Across RQ1–RQ3, participant quotes reveal a consistent progression from reflective awareness to action-oriented challenges. Participants described gains in self-efficacy, emotional clarity, and planning awareness (RQ1), while expressing distinct preferences for either free-form flexibility or structured guidance in shaping how reflection unfolded (RQ2). Voice journaling was generally perceived as convenient and emotionally expressive, supporting spontaneity and disclosure, but also introduced environmental and motivational constraints that affected sustained use (RQ3). Taken together, these findings suggest that while reflective systems can support insight and perceived progress, participants’ experiences are shaped by an interplay between structure, modality, and everyday contextual constraints, which in turn influences their ability to translate reflection into sustained change.}

\begin{table*}[!htbp]
\centering
\caption{Representative participant quotes for RQ1 (User perception on adaptive coping and regulatory behavior-emotions).}
\label{tab:quotes-rq1}
\small
\setlength{\tabcolsep}{4pt}
\renewcommand{\arraystretch}{1.22}

\begin{tabular}{@{}p{2.6cm}p{0.60\textwidth}p{1.8cm}p{2.4cm}@{}}
\toprule
\textbf{Code} & \textbf{Quote} & \textbf{ID} & \textbf{Condition} \\
\midrule
Self-Efficacy & \emph{“I definitely feel more in control… it showed me there are things I can do.”} & P72661 & Gross-guided \\
 & \emph{“It gave me a sense of direction instead of feeling lost.”} & P53782 & Gross-guided\\
 & \emph{“When I said it out loud, it made me believe I could handle it better.”} & P91592 & Gross-guided \\
\midrule
Obstacle Identification & \emph{“I did not like coming up with alternative solutions… it was the same list I kept writing.”} & P75096 & Free-form  \\
 & \emph{“I realized my recordings were basically the same… that helped me see the pattern.”} & P37445 & Free-form  \\
 & \emph{“I kept noticing the same triggers coming up again and again.”} & P69882 & Gross-guided \\
\midrule
Strategy Planning & \emph{“It allowed me to plan for tomorrow, instead of just thinking about regrets.”} & P53782 & Gross-guided \\
 & \emph{“The exercise made me identify barriers and alternative ways.”} & P37445 & Free-form \\
 & \emph{“I started thinking about actual steps, not just feelings.”} & P81956 & Free-form  \\
\midrule
Emotion Regulation & \emph{“The prompts made me reflect instead of blaming myself, and I saw I could reframe.”} & P69882 & Gross-guided \\
 & \emph{“I became more aware of how I react emotionally.”} & P91592 & Gross-guided \\
 & \emph{“It showed me that even if I can’t change everything, I can change how I respond.”} & P60981 & Gross-guided \\
\bottomrule
\end{tabular}
\end{table*}

\begin{table*}[!htbp]
\centering
\caption{Representative participant quotes for RQ2 (User perceptions of Gross-guided vs. Free-form).}
\label{tab:quotes-rq2}
\small
\renewcommand{\arraystretch}{1.22}

\begin{tabular}{@{}p{2.6cm}p{0.60\textwidth}p{1.8cm}p{2.4cm}@{}}
\toprule
\textbf{Subcode} & \textbf{Quote} & \textbf{ID} & \textbf{Condition} \\
\midrule
Freedom / Flexibility
& \emph{“I just told the story in my own way, which felt natural.”} 
& P37445 & Free-form \\
& \emph{“It was flexible, I could go in any direction.”} 
& P42520 & Free-form \\
& \emph{“Sometimes I wandered, but that helped me discover things.”} 
& P75096 & Free-form \\
& \emph{“I liked being able to just say whatever was on my mind.”} 
& P81956 & Free-form \\
\midrule
Structured Guidance
& \emph{“The structure made me think through steps instead of just venting.”} 
& P37444 & Gross-guided \\
& \emph{“The prompts reminded me to reframe instead of staying stuck.”} 
& P60981 & Gross-guided \\
& \emph{“It pushed me to recognize patterns in my reactions.”} 
& P69355 & Gross-guided \\
& \emph{“It gave me clarity about what part I could actually control.”} 
& P72661 & Gross-guided \\
\bottomrule
\end{tabular}
\end{table*}

\begin{table*}[!htbp]
\centering
\caption{Representative participant quotes for RQ3 (User perception on voice journaling).}
\label{tab:quotes-voice}
\small
\renewcommand{\arraystretch}{1.22}

\begin{tabular}{@{}p{2.6cm}p{0.60\textwidth}p{1.8cm}p{2.4cm}@{}}
\toprule
\textbf{Subcode} & \textbf{Quote} & \textbf{ID} & \textbf{Condition} \\
\midrule
Convenience 
& \emph{“It’s easily accessible … I could be on the bus … I’m more likely to voice record …”} 
& P37445 & Free-form \\
& \emph{“I liked phone-based recording for privacy and convenience …”} 
& P42520 & Free-form \\
& \emph{“It was better than I expected … the user experience was really good.”} 
& P60981 & Gross-guided \\
& \emph{“I loved that it was way faster than writing it down …”} 
& P37444 & Gross-guided \\
\midrule
Emotional Clarity 
& \emph{“Speaking helped me realize what I was really feeling.”} 
& P81956 & Free-form \\
& \emph{“It brought out emotions I didn’t know were there.”} 
& P26409 & Free-form \\
& \emph{“Saying things aloud gave me emotional clarity.”} 
& P69882 & Gross-guided \\
& \emph{“The voice entries helped me process better than writing.”} 
& P91592 & Gross-guided \\
\midrule
Challenges 
& \emph{“It was hard to find a quiet, private space …”} 
& P26409 & Free-form \\
& \emph{“I dislike that you cannot really keep track visually … sometimes I repeated myself …”} 
& P81956 & Free-form \\
& \emph{“You have to find a place to do it.”} 
& P91592 & Gross-guided \\
& \emph{“Sometimes I wouldn’t voice journal outside of that room.”} 
& P68786 & Gross-guided \\
\bottomrule
\end{tabular}
\end{table*}

\begin{table*}[!htbp]
\centering
\caption{Representative participant quotes related to challenges in self-reflection for change.}
\label{tab:quotes-challenges}
\small
\renewcommand{\arraystretch}{1.22}

\begin{tabular}{@{}p{2.6cm}p{0.60\textwidth}p{1.8cm}p{2.4cm}@{}}
\toprule
\textbf{Subcode} & \textbf{Quote} & \textbf{ID} & \textbf{Condition} \\
\midrule
Stress / Overwhelm 
& \emph{“Sometimes I just didn’t have the energy to reflect after work, even though I knew it might help.”} 
& P37444 & Gross-guided \\
& \emph{“The hardest part was actually remembering to do it in the middle of my busy day.”} 
& P53782 & Gross-guided \\
& \emph{“I found it difficult to be honest at times, knowing someone might read or hear my recordings.”} 
& P26409 & Free-form \\
\midrule
Journaling Environment 
& \emph{“Finding a quiet space to talk was a challenge. I live with family, so I felt self-conscious.”} 
& P68786 & Gross-guided \\
& \emph{“I often repeated myself… it felt like I wasn’t moving forward.”} 
& P81956 & Free-form \\
\bottomrule
\end{tabular}
\end{table*}

\end{document}